\documentclass{article} 
\usepackage{iclr2027_conference,times}

\usepackage{amsmath,amsfonts,bm}

\def\eqref#1{equation~\ref{#1}}

\def\1{\bm{1}}

\DeclareMathAlphabet{\mathsfit}{\encodingdefault}{\sfdefault}{m}{sl}
\SetMathAlphabet{\mathsfit}{bold}{\encodingdefault}{\sfdefault}{bx}{n}

\usepackage{hyperref}
\usepackage{url}

\usepackage{booktabs} 
\usepackage{graphicx}
\usepackage{amsmath} 
\usepackage{multirow}
\usepackage{algorithm}
\usepackage{algpseudocode}
\usepackage[table]{xcolor}
\usepackage{amssymb} 
\newcommand{\oodcell}[1]{\cellcolor{gray!12}#1}

\title{Learning Transferable Reaction Mechanisms from Visual Chemical Knowledge}

\author{
\textbf{Yujian Yuan}$^{1}$, 
\textbf{Jiaxin Xu}$^{1}$,
\textbf{Xin Cai}$^{2}$, 
\textbf{Yufan Chen}$^{1}$, 
\textbf{Zhichao Tan}$^{1}$,
\textbf{Ziqi Zhou}$^{3}$, 
\textbf{Hanyu Gao}$^{1,*}$
\\
$^{1}$The Hong Kong University of Science and Technology
\quad
$^{2}$The Chinese University of Hong Kong
\\
$^{3}$ University of Edinburgh
\\
\texttt{yyuanbn@connect.ust.hk, hanyugao@ust.hk, *:Corresponding author}
}

\iclrfinalcopy 
\begin{document}

\maketitle

\begin{abstract}

Reaction mechanisms describe the step-by-step transformations underlying
chemical reactions and are central to reaction analysis and synthesis.
Learning-based models have achieved strong performance on established
mechanism-prediction benchmarks, but transferring them to unseen chemistry
remains challenging.
Such transfer is difficult because familiar mechanisms must be applied
to unfamiliar molecular structures, and some target mechanisms may be poorly
covered by the training data.
To address these challenges, we introduce \textbf{MechaVLM}, a visual framework
that combines transferable chemical representations with external mechanistic
knowledge.
It learns reusable visual features through multiscale chemical grounding and
cross-rendering contrastive learning.
For open-book prediction, MechaVLM retrieves a fixed set of precedents from
70,384 literature mechanism figures and re-reads relevant visual evidence as
the molecular state evolves, directly using the figures without symbolic
mechanism parsing.
An atom-indexed language decoder then recursively generates executable electron
edits to construct the complete mechanism.
We further introduce \textbf{MechBench}, a challenging literature-derived benchmark with
2,184 mechanisms and 9,146 elementary steps.
Across cross-dataset and literature-derived benchmarks, MechaVLM establishes
strong zero-shot mechanism prediction.
Its closed-book model alone improves Step/Pathway Top-1 by 12.50/13.93
on FlowER$\rightarrow$ReactMech, while external visual precedents unlock further
gains on challenging OOD reactions.
The learned representation also generalizes beyond mechanism prediction to atom
mapping and reaction center prediction. 
Codes and benchmarks will be released.

\end{abstract}

\section{Introduction}

Understanding how chemical reactions proceed through elementary transformations is central to reaction analysis and synthesis.
Recent learning-based models have achieved strong  reaction mechanism prediction
performance on large mechanistically annotated datasets~\citep{joung2025electron,das2026deepmech}.
However, transferring these models to a different reaction distribution beyond the training distribution remains challenging~\citep{dang2026learning}. 
Because obtaining mechanism annotations for every new domain is costly,
repeatedly adapting a model to new chemistry is impractical.
We therefore study \emph{transferable reaction mechanism prediction}, where a
model trained on existing mechanism data must generalize zero-shot to unseen
chemical domains.

Such transfer can be difficult for two reasons.
First, a familiar mechanism may need to be applied to molecular structures that
differ substantially from those seen during training.
This requires representations that retain reusable local chemical information.
Second, the relevant mechanism may be poorly covered by the training data, so
the model must rely on knowledge beyond its learned parameters.
To address these, we propose using \textbf{visual modeling} as a natural way to support both forms of transfer: (1)
First, molecular diagrams expose atoms, bonds, and local structures directly,
providing a representation space in which reusable chemical patterns can be
learned across different reactions.
(2) Second, the same visual space provides direct access to external mechanistic knowledge.
Chemical literature contains abundant expert-curated mechanism figures showing molecular structures, intermediates, and electron-flow arrows across diverse real-world reactions~\citep{dang2026learning}.
Using such figures in a symbolic pipeline typically requires first extracting
their molecular structures and mechanistic information, which can be costly,
error-prone, and may discard useful visual context~\citep{qian2023molscribe}.
In contrast, when both the input reaction and external precedents are
represented as images, retrieval can be performed directly in the same visual
space.
This avoids both prerequisite symbolic extraction and an additional cross-modal retrieval alignment stage, while preserving the original information in the
literature figures.

Retrieving a relevant mechanism figure is not enough.
A literature figure usually shows a complete multi-step pathway, while the
model predicts one elementary step at a time.
Only part of the retrieved figure may be useful for the current step, especially
when the precedent and target reaction differ in structure.
As the molecular state evolves, the useful part may also change.
The key challenge is to transfer the right local mechanistic evidence to each
prediction step.

\begin{figure}[t]
    \centering
    \includegraphics[width=1.0\textwidth]{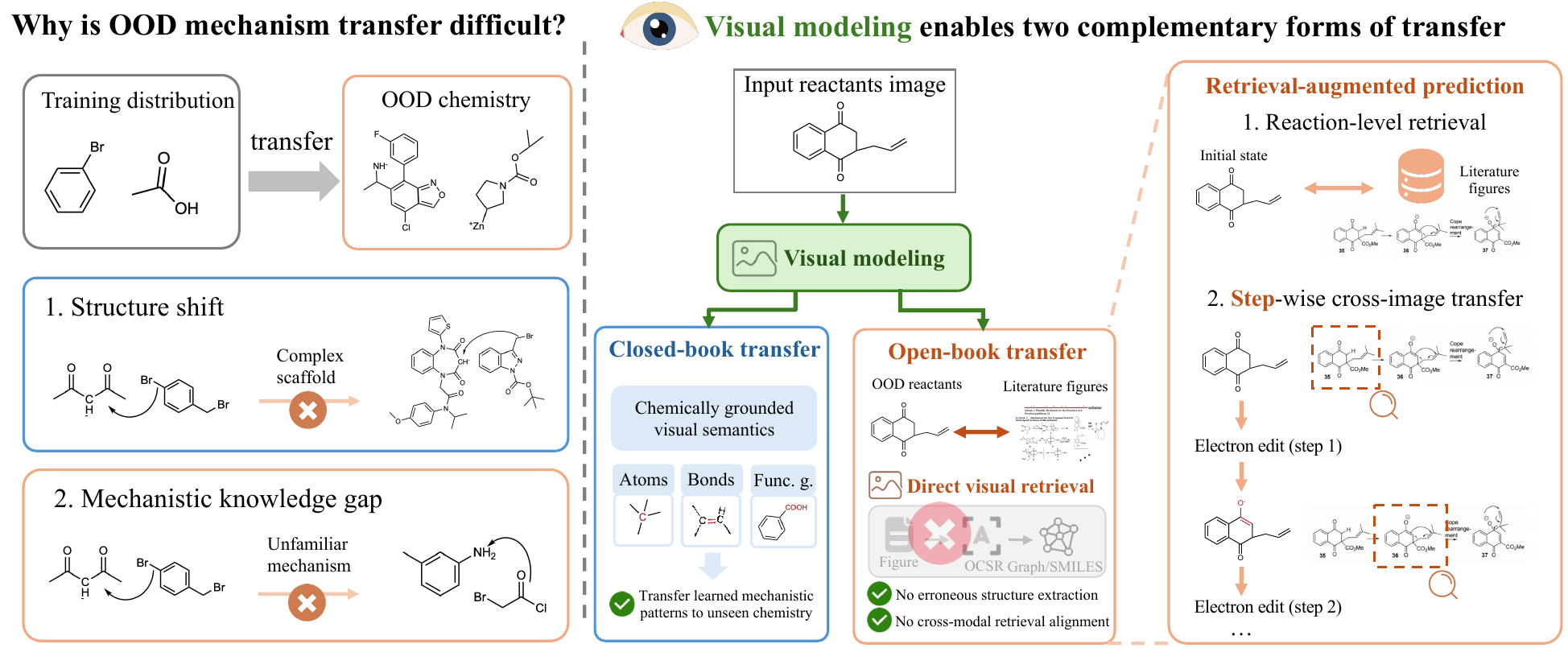}
    \vspace{-2.5em}
    \caption{Transferable reaction mechanism prediction with MechaVLM. Visual modeling supports complementary closed-book and open-book transfer for OOD chemistry, while retrieved pathway-level precedents are dynamically transferred to step-wise electron-edit predictions.
    }
    \label{fig:overview}
    \vspace{-1.5em}
\end{figure}

To address these challenges, we introduce \textbf{MechaVLM}, a visual
framework for transferable reaction mechanism prediction.
MechaVLM combines representation learning with retrieval-based mechanism
transfer in a common visual space.
It first learns chemically grounded representations through multiscale chemical
supervision and cross-rendering contrastive learning, encouraging reusable
local chemical structure to remain stable across reaction distributions.
When external knowledge is available, MechaVLM retrieves a fixed set of visual
precedents using the initial reactant image and introduces step-wise cross-image
transfer to re-read their local evidence as the molecular state evolves.
The resulting evidence is decoded into atom-indexed electron edits, which are
applied recursively to generate the complete mechanism.

To evaluate transfer to challenging mechanisms, we introduce
\textbf{MechBench}, a literature-derived benchmark containing
2,184 mechanisms and 9,146 elementary steps.
Across cross-dataset and literature-derived benchmarks, MechaVLM consistently
improves zero-shot mechanism prediction.
In FlowER$\rightarrow$ReactMech transfer, the closed-book model improves Step
and Pathway Top-1 accuracy over the strongest baselines by 12.50 and 13.93.
Visual precedents provide further gains in the open-book setting and outperform alternative retrieval-augmented baselines. The
learned representation also transfers effectively to reaction-related tasks (atom mapping \& reaction
center prediction).

Our primary contributions are summarized as follows:
\begin{itemize}
\vspace{-0.5em}
\item We study transferable reaction mechanism prediction, focusing on
zero-shot transfer across reaction distributions, and introduce
\textbf{MechBench}, a literature-derived benchmark with 2,184 mechanisms and
9,146 elementary steps for evaluating transfer to unseen chemistry.

\item We introduce \textbf{MechaVLM}, a visual framework that combines
transferable chemical representation learning with retrieval-augmented
mechanism prediction.
It learns reusable local  structure through multiscale grounding and
cross-rendering contrastive learning, and uses step-wise cross-image transfer
to adapt precedent evidence as the molecular state evolves.

\item Extensive experiments show that MechaVLM achieves state-of-the-art zero-shot
transfer across cross-dataset and literature-derived mechanism benchmarks.
The learned visual representation further transfers to atom mapping and reaction
center prediction, while access to external visual precedents provides substantial
additional gains over closed-book results.

\end{itemize}

\begin{figure}[t]
    \centering
    \includegraphics[width=1.0\textwidth]{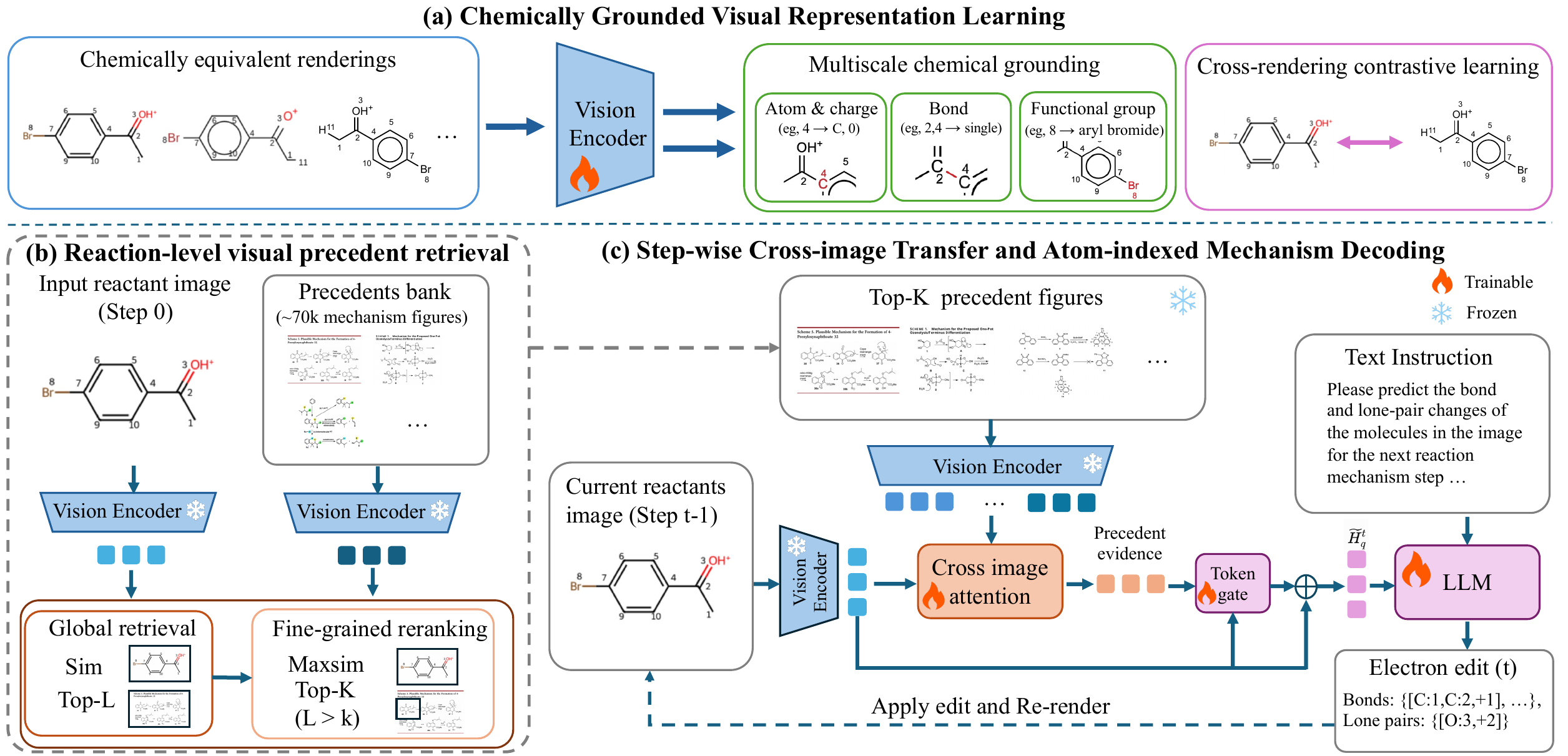}
    \vspace{-2em}
    \caption{
Overview of \textbf{MechaVLM}.
(a) Chemically grounded and rendering-robust visual representation learning.
(b) Reaction-level precedent retrieval via global screening and fine-grained reranking.
(c) Step-wise cross-image transfer dynamically extracts relevant evidence from the retrieved precedents and grounds it to indexed query atoms for executable electron-edit decoding. The updated state is re-rendered for the next step.
}
    \label{fig:overview}
\end{figure}

\section{Related Work}

\textbf{Reaction Mechanism Prediction.}
Reaction mechanism prediction models chemical reactions as sequences of
elementary electron movements and molecular state transitions.
Early approaches predicted electron-transfer paths for textbook organic
reactions~\citep{kayala2012reaction}, followed by learning-based methods that
model electron flow through graph edits or electron redistribution
~\citep{bradshaw2018generative,bi2021non}.
More recently, FlowER~\citep{joung2025electron} models continuous electron
redistribution with Bond--Electron matrices, while DeepMech
~\citep{das2026deepmech} further evaluates generalization across reaction
classes and chemical components.
Despite strong performance on established benchmarks, these methods degrade
substantially when transferred to chemistry beyond the training distribution
~\citep{dang2026learning}.
We address this limitation from a visual perspective to learn transferable
chemical representations.

\textbf{Chemical Vision-Language Learning and Visual Retrieval.}
Our work is related to two lines of visual modeling.
Recent chemical VLMs, including ChemVLM~\citep{li2025chemvlm},
TinyChemVL~\citep{zhao2026tinychemvl}, and
ChemVLR~\citep{zhao2026chemvlr}, have demonstrated that molecular and reaction
images can encode rich chemical information for understanding, reasoning, and
prediction.
However, these models are not designed specifically for transferring reaction
mechanisms across different chemical distributions.
A complementary line of work explores retrieval directly from visual content:
ColPali~\citep{faysse2025colpali} and VisRAG~\citep{yu2025visrag}, for example,
preserve document information in image form rather than relying entirely on
intermediate textual representations.
In contrast, our MechaVLM performs image-to-image retrieval from reactant images
to literature mechanism figures, and further transfers the retrieved evidence
to step-wise mechanism prediction.

\section{Method}

We propose MechaVLM, a visual framework for transferable reaction mechanism
prediction.
MechaVLM combines two complementary sources of transfer.
First, it learns chemically grounded visual representations through multiscale
chemical supervision and cross-rendering contrastive learning, improving
generalization across reaction distributions.
Second, in the open-book setting, it retrieves mechanism figures directly as
visual precedents and reuses a fixed set throughout the reaction, while
step-wise cross-image transfer selects the relevant evidence as the molecular
state evolves.
Finally, an atom-indexed language decoder generates executable electron edits,
which are applied and re-rendered recursively to produce the complete mechanism.

\subsection{Problem Formulation: Transferable Reaction Mechanism Prediction}
\label{sec:problem}

\textbf{Reaction mechanism prediction.}
Given a reactant set $R$, let $S^0$ denote its initial molecular state.
We represent a reaction mechanism as a sequence of elementary electron edits~\citep{joung2025electron}
\begin{equation}
Y
=
\left(
\Delta^1,\ldots,\Delta^T
\right),
\qquad
S^t
=
S^{t-1}
\oplus
\Delta^t,
\end{equation}
where each $\Delta^t$ denotes bond and lone-pair electron changes,
and $\oplus$ applies the edit to reach the next molecular state.
Starting from the initial reactant state $S^0$ with no ground-truth products provided, the predictor recursively generates elementary electron
edits until termination.

\textbf{Transfer setting.}
We focus on zero-shot transfer across reaction distributions.
A model is trained on source reactions
$\mathcal{D}_{\mathrm{src}}\sim P_{\mathrm{src}}(R,Y)$
and evaluated on target reactions
$\mathcal{D}_{\mathrm{tgt}}\sim P_{\mathrm{tgt}}(R,Y)$,
where $P_{\mathrm{src}}\neq P_{\mathrm{tgt}}$.
No target-domain reactions are used for fine-tuning.
The distribution shift may involve molecular scaffolds, reaction families, or
broader chemical contexts.
We evaluate two inference settings:
(1) in the \emph{closed-book} setting, prediction relies only on the knowledge
encoded in the model parameters.
(2) in the \emph{open-book} setting, the model may additionally leverage a fixed
external knowledge bank $\mathcal{B}$.
This distinction studies two complementary forms of transfer:
generalization from learned chemical representations and transfer from external mechanistic knowledge.

\subsection{Chemically Grounded Visual Representation Learning}
\label{sec:visual_grounding}

Transfer across reaction distributions requires visual representations that
capture chemical structure rather than drawing-specific patterns.
We therefore pretrain the visual encoder with two complementary objectives:
\emph{multiscale chemical grounding}, which teaches the model to recognize
local chemical units, and \emph{cross-rendering contrastive learning}, which makes the
representation stable across different renderings of the same molecule.
Given a rendered reactant image
$I=\rho_{\nu}(R)$,
the visual encoder $\mathcal{E}_{\theta}$ produces dense features
$
    H
    =
    \mathcal{E}_{\theta}(I)
    =
    \{h_1,\ldots,h_L\}.
$

\textbf{Multiscale chemical grounding.}
We ground visual features in explicit chemical semantics at three levels:
atoms, bonds, and functional groups.
Specifically, the model predicts atom identity and formal charge, bond
connectivity and type, and local functional-group context.
The corresponding supervision is automatically derived from the reactant
structure used to render each image.
To identify the target chemical units, we render atom indices directly in the
image.
To specify the atoms involved in each auxiliary task, we overlay atom indices beside their corresponding atom in the image.
These indices are randomly permuted across renderings.
Let
$
    \mathcal{T}_{\mathrm{chem}}
    =
    \{
        \mathrm{atom},
        \mathrm{bond},
        \mathrm{functional\text{-}group}
    \}.
$
For task $k\in\mathcal{T}_{\mathrm{chem}}$, queried unit $u$, target $y$, and
instruction $\pi_k$, we optimize
\begin{equation}
    \mathcal{L}_{\mathrm{chem}}
    =
    -\mathbb{E}_{(R,k,u,y),\,\nu}
    \left[
        \log
        P_{\theta}
        \left(
            y
            \mid
            \rho_{\nu}(R),u,\pi_k
        \right)
    \right],
    \qquad
    \nu\sim\mathcal{G}_{\mathrm{vis}}.
\end{equation}
By explicitly supervising these local chemical units, the encoder is encouraged
to capture reusable chemical structure rather than relying on dataset-specific
visual patterns.

\textbf{Cross-rendering contrastive learning.}
The same chemical structure can appear in substantially different visual forms
under different rendering conditions.
We therefore generate multiple chemistry-preserving renderings of the same
reactants and align their representations through contrastive learning.
For each reactant structure $R_i$, we sample two renderings
\[
    I_i^{(a)}=\rho_{\nu_a}(R_i),
    \qquad
    I_i^{(b)}=\rho_{\nu_b}(R_i),
\]
where $\nu_a,\nu_b\sim\mathcal{G}_{\mathrm{vis}}$ vary molecular layout,
drawing style, resolution, cropping, atom indexing, and additional graphical
perturbations while preserving the underlying chemical structure.
For each rendering, we obtain a normalized global representation of the vision encoder $\mathcal{E}_{\theta}$:
\begin{equation}
    z(I)
    =
    \operatorname{Norm}
    \left(
        \operatorname{Pool}
        \left(
            \mathcal{E}_{\theta}(I)
        \right)
    \right).
\end{equation}
We align representations of the same reactants using a symmetric in-batch
contrastive objective,
\begin{equation}
    \mathcal{L}_{\mathrm{ctr}}
    =
    \frac{1}{2}
    \left[
        \operatorname{InfoNCE}
        \left(
            Z^{(a)},Z^{(b)}
        \right)
        +
        \operatorname{InfoNCE}
        \left(
            Z^{(b)},Z^{(a)}
        \right)
    \right],
\end{equation}
with cosine similarity and temperature $\tau$.

Together with multiscale chemical grounding, the representation objective is
\begin{equation}
    \mathcal{L}_{\mathrm{repr}}
    =
    \mathcal{L}_{\mathrm{chem}}
    +
    \lambda_{\mathrm{ctr}}
    \mathcal{L}_{\mathrm{ctr}}.
\end{equation}

\subsection{Retrieval-Augmented Prediction with Visual Precedents}
\label{sec:retrieval}

In the open-book setting, MechaVLM augments mechanism prediction with
mechanism figures retrieved from an external precedent bank.
MechaVLM retrieves a fixed set of precedents from the initial reactants and
reuses them throughout the mechanism rollout.
At each prediction step, it re-reads these precedents against the current
molecular state, allowing different regions to contribute as the mechanism
evolves.

\textbf{Reaction-level precedent retrieval.}
We construct an external precedent bank
$\mathcal{B}=\{I_i^p\}_{i=1}^{M}$ containing $M$ mechanism figures.
Each figure is kept in its original visual form, including molecular
structures, intermediates, electron-flow arrows, and textual annotations. Given the initial reactant image
$I_q^{\mathrm{init}}=\rho(S_q^0)$,
we retrieve precedents using the frozen encoder
$\mathcal{E}_{\theta_0}$ from Sec.~\ref{sec:visual_grounding}.
We first use the global representation to shortlist the top-$L$ candidates:
\begin{equation}
\begin{aligned}
    s_i^{\mathrm{g}}
    &=
    \operatorname{sim}
    \left(
        z(I_q^{\mathrm{init}}),
        z(I_i^p)
    \right),
    \qquad
    \mathcal{C}_L
    =
    \operatorname{TopL}_{I_i^p\in\mathcal{B}}
    s_i^{\mathrm{g}}.
\end{aligned}
\end{equation}
Global similarity may miss a useful precedent when the query reactants match
only a small region of a complete mechanism figure.
We therefore rerank the shortlisted candidates using late interaction~\citep{faysse2025colpali} between
their dense visual features.
Let
$H_q^{\mathrm{init}}
=
\{h_{q,m}^{\mathrm{init}}\}_{m=1}^{N_q}$
and
$H_i^p
=
\{h_{i,n}^p\}_{n=1}^{N_i}$
denote the $N_q$ query and $N_i$ precedent visual features, respectively.
We compute
\begin{equation}
\begin{aligned}
    s_i^{\mathrm{LI}}
    &=
    \frac{1}{N_q}
    \sum_{m=1}^{N_q}
    \max_{1\leq n\leq N_i}
    \operatorname{sim}
    \left(
        h_{q,m}^{\mathrm{init}},
        h_{i,n}^p
    \right), \\
    \mathcal{R}_K
    &=
    \operatorname{TopK}_{I_i^p\in\mathcal{C}_L}
    s_i^{\mathrm{LI}}.
\end{aligned}
\end{equation}
where $s_i^{\mathrm{LI}}$ measures the local visual similarity between the
query and precedent $i$.
This allows different parts of the reactant image to match relevant regions
within a complete mechanism figure.
The resulting top-$K$ precedent set $\mathcal{R}_K$ is kept fixed throughout
the mechanism rollout.

\textbf{Step-wise cross-image transfer.}
Although the retrieved precedents remain fixed, the useful information within
them can change as the molecular state evolves.
At prediction step $t$, we render and encode the current state:
\begin{equation}
    I_q^t
    =
    \rho(S_q^{t-1}),
    \qquad
    H_q^t
    =
    \mathcal{E}_{\theta_0}(I_q^t).
\end{equation}
We concatenate the dense features of the retrieved precedents and use cross-image attention to read precedent features relevant to the
current molecular state:
\begin{equation}
    H_{\mathcal{R}}
    =
    \operatorname{Concat}_{I_i^p\in\mathcal{R}_K}
    H_i^p,
\end{equation}
\begin{equation}
    O_{\mathrm{ext}}^t
    =
    \operatorname{softmax}
    \left(
        \frac{
            (H_q^t W_Q)
            (H_{\mathcal{R}} W_K)^\top
        }{
            \sqrt{d}
        }
    \right)
    (H_{\mathcal{R}} W_V),
\end{equation}
where $W_Q$, $W_K$, and $W_V$ are learned projections and $d$ is the
attention dimension.
Because $H_q^t$ changes after each predicted edit, this attention is
recomputed at every step.
The model can therefore draw on different regions of the same precedent
figures as the mechanism progresses.

Finally, as not every matched precedent region is useful for every query feature, we use a token-wise gate to control how much retrieved information is
introduced into each query feature:
\begin{equation}
\begin{aligned}
    G^t
    &=
    \sigma
    \left(
        W_G
        \left[
            H_q^t;
            O_{\mathrm{ext}}^t W_O
        \right]
        +
        b_G
    \right), \\
    \widetilde{H}_q^t
    &=
    H_q^t
    +
    G^t
    \odot
    \left(
        O_{\mathrm{ext}}^t W_O
    \right),
\end{aligned}
\end{equation}
where $\sigma$ is the sigmoid function and $\odot$ denotes element-wise
multiplication.
The resulting representation $\widetilde{H}_q^t$ is used to predict the next
elementary edit in Sec.~\ref{sec:mechanism_decoding}.

\subsection{Executable Mechanism Decoding}
\label{sec:mechanism_decoding}

At each mechanism step, MechaVLM predicts an executable electron edit on the
current molecular state.
The same decoder is used in both inference settings: it operates on
$H_q^t$ in the closed-book setting and on the precedent-enhanced
representation $\widetilde{H}_q^t$ from Sec.~\ref{sec:retrieval} in the
open-book setting.

\textbf{Atom-indexed electron-edit decoding.}
At step $t$, the current molecular state is rendered with atom indices, allowing
the model to refer directly to specific atoms in the predicted transformation.
The visual representation $\widetilde{H}_q^t$ is provided to the language
decoder of the VLM, which autoregressively generates the structured electron
edit:
\begin{equation}
    \Delta_q^t
    \sim
    P_{\phi}
    \left(
        \cdot
        \mid
        \widetilde{H}_q^t
    \right),
\end{equation}
where $\widetilde{H}_q^t=H_q^t$ in closed-book and denotes the
precedent-enhanced representation in open-book.

Each elementary step consists of bond and lone-pair electron changes:
\begin{equation}
\begin{aligned}
    \Delta_q^t
    &=
    \left(
        \Delta_{q,\mathrm{bond}}^t,
        \Delta_{q,\mathrm{lp}}^t
    \right), \\
    \Delta_{q,\mathrm{bond}}^t
    &=
    \left\{
        \left(
            a_i,a_j,\delta_{ij}
        \right)
    \right\},
    \qquad
    \Delta_{q,\mathrm{lp}}^t
    =
    \left\{
        \left(
            a_i,\delta_i^e
        \right)
    \right\},
\end{aligned}
\end{equation}
where $a_i$ denotes an indexed atom,
$\delta_{ij}$ specifies a bond change between atoms $a_i$ and $a_j$,
and $\delta_i^e$ specifies a lone-pair electron change on atom $a_i$.
The predicted edit can therefore be directly applied to the current molecular
state.

\textbf{Iterative mechanism rollout.}
After predicting $\widehat{\Delta}_q^t$, we apply the edit and re-render the
updated :
\begin{equation}
    S_q^t
    =
    S_q^{t-1}
    \oplus
    \widehat{\Delta}_q^t,
    \qquad
    I_q^{t+1}
    =
    \rho(S_q^t).
\end{equation}
The updated image is then used to predict the next elementary step.
In the open-book setting, it also re-conditions the step-wise cross-image
transfer in Sec.~\ref{sec:retrieval}, while the retrieved precedent set
$\mathcal{R}_K$ remains fixed.
This process continues until a non-electron-movement action is predicted.

\textbf{Training.}
After representation training, we freeze the vision encoder and fine-tune
the language decoder on source-domain mechanism data. The open-book
model additionally trains the cross-image transfer and fusion modules.
At step $t$, the ground-truth state $S_q^{t-1}$ is rendered as input, and the
language decoder is trained with teacher forcing to generate
the serialized electron edit $\Delta_q^t$ using the standard LLM loss.
For open-book training, precedents are retrieved once from initial
reactants and reused throughout the trajectory, while cross-image transfer is
recomputed for each state.

\section{Experiments}
\vspace{-1em}

\textbf{Datasets.}
We evaluate on four reaction-mechanism datasets.
FlowER~\citep{joung2025electron} contains
250,782/2,801/28,049 train/validation/test reactions
(1.45M/15.7K/162.0K elementary steps).
ReactMech~\citep{das2026deepmech} contains 29,604 mechanism pathways and
104,964 elementary steps across 67 reaction classes.
For zero-shot benchmarks, FukuyamaBench~\citep{dang2026learning} contains 319 pathways
with 1,997 elementary steps from Fukuyama Mechanism Book. MechBench is below.

\textbf{MechBench} is our proposed literature-derived benchmark
constructed from reaction-mechanism figures in 417 publications from
\emph{JACS}, \emph{JOC}, and \emph{Organic Letters}.
After manually screening articles and retaining only complete, self-contained
mechanistic pathways, we curate 2,184 pathways comprising
9,146 elementary steps, with all reconstructed mechanisms verified
against the original figures by three chemical experts.
Data construction details are provided in the Appendix.

\textbf{External visual precedent bank.}
We construct a fixed bank
$\mathcal{B}$ of 70,384 reaction mechanism figures collected from the chemical literature.
Figures are retained in their original visual form, including molecular
structures, mechanistic arrows, and are indexed
without conversion to symbolic molecular representations.
The bank is used for open-book precedent retrieval during training and
inference, without structured mechanism labels or query--precedent relevance
annotations.
We remove overlaps with all training and benchmark reactions to prevent retrieval leakage.
Details are in Appendix.

\textbf{Metrics.}
Following~\citep{joung2025electron}, we report \emph{Top-$k$ step accuracy} and
\emph{Top-$k$ pathway accuracy}, with $k\in\{1,5\}$.
Step accuracy measures whether the ground-truth next intermediate is recovered
among the top-$k$ predictions for an elementary step.
Pathway accuracy requires the complete sequence of mechanistic
intermediates to be recovered within the corresponding beam.

\textbf{Implementation.}
MechaVLM is initialized from Qwen2.5-VL-3B~\citep{bai2025qwen25vltechnicalreport}.
In the first stage, we fully fine-tune the vision encoder with
$\mathcal{L}_{\mathrm{repr}}$ for 2 epochs while keeping the LLM
frozen, using AdamW with a learning rate of $1\times10^{-4}$ and weight decay
$0.1$.
We then freeze the vision encoder and use it to encode the external precedent
bank once, building a nearest-neighbor index over the resulting global
representations.
In the second stage, we perform mechanism SFT by fully fine-tuning the LLM with a lr of $3\times10^{-6}$.
Both stages use a batch size of 32 on 8 NVIDIA H800.

\begin{table*}[t]
\centering
\footnotesize
\caption{
Cross-domain reaction mechanism prediction (\%).
Models are trained on the source dataset indicated in the first column.
White cells report in-distribution (ID) performance, while gray cells denote
zero-shot out-of-distribution (OOD) transfer without target-domain adaptation.
}
\label{tab:cross_domain}
\setlength{\tabcolsep}{10pt}
\renewcommand{\arraystretch}{1.08}

\resizebox{\textwidth}{!}{
\begin{tabular}{@{}llcccccccc@{}}
\toprule
&
&
\multicolumn{4}{c}{\textbf{FlowER}}
&
\multicolumn{4}{c}{\textbf{ReactMech}} \\
\cmidrule(lr){3-6}
\cmidrule(lr){7-10}

&
&
\multicolumn{2}{c}{Step}
&
\multicolumn{2}{c}{Pathway}
&
\multicolumn{2}{c}{Step}
&
\multicolumn{2}{c}{Pathway} \\
\cmidrule(lr){3-4}
\cmidrule(lr){5-6}
\cmidrule(lr){7-8}
\cmidrule(lr){9-10}

\textbf{Source}
& \textbf{Method}
& Top-1 & Top-5
& Top-1 & Top-5
& Top-1 & Top-5
& Top-1 & Top-5 \\
\midrule

\multirow{9}{*}{FlowER}

& NERF~\citep{bi2021non}
& 67.11 & 72.60
& 65.51 & 75.58
& \oodcell{33.32} & \oodcell{43.12}
& \oodcell{14.93} & \oodcell{36.32} \\

& Graph2SMILES~\citep{tu2022permutation}
& 89.09 & 98.66
& 92.51 & 97.76
& \oodcell{42.71} & \oodcell{55.00}
& \oodcell{31.84} & \oodcell{61.19} \\

& Graph2SMILES+H~\citep{tu2022permutation}
& 87.39 & 97.67
& 89.22 & 96.19
& \oodcell{40.42} & \oodcell{52.37}
& \oodcell{29.67} & \oodcell{59.74} \\

& MT~\citep{schwaller2019molecular}
& 88.75 & 98.92
& 88.31 & 97.59
& \oodcell{44.52} & \oodcell{52.45}
& \oodcell{\underline{39.30}} & \oodcell{\underline{58.21}} \\

& FlowER~\citep{joung2025electron}
& 88.48 & 98.60
& 88.97 & 97.35
& \oodcell{\underline{56.84}} & \oodcell{\underline{63.18}}
& \oodcell{34.58} & \oodcell{51.47} \\

& ChemVLM-8B~\citep{li2025chemvlm}
& 86.73 & 96.52
& 86.94 & 95.72
& \oodcell{31.42} & \oodcell{39.08}
& \oodcell{12.26} & \oodcell{16.84} \\

& TinyChemVL~\citep{zhao2026tinychemvl}
& 88.21 & 98.34
& 91.06 & 97.89
& \oodcell{37.84} & \oodcell{45.26}
& \oodcell{16.05} & \oodcell{22.71} \\

& ChemVLR-8B~\citep{zhao2026chemvlr}
& 88.54 & 98.61
& 91.87 & 98.32
& \oodcell{41.56} & \oodcell{48.73}
& \oodcell{19.83} & \oodcell{27.16} \\

& \textbf{MechaVLM (ours)}
& 90.02 & 99.20
& 93.89 & 99.14
& \oodcell{\textbf{69.34}} & \oodcell{\textbf{78.42}}
& \oodcell{\textbf{53.23}} & \oodcell{\textbf{68.71}} \\

\midrule

\multirow{7}{*}{ReactMech}

& NERF~\citep{bi2021non}
& \oodcell{8.62} & \oodcell{12.61}
& \oodcell{0.04} & \oodcell{0.12}
& 60.58 & 60.84
& 47.76 & 47.76 \\

& Graph2SMILES~\citep{tu2022permutation}
& \oodcell{3.03} & \oodcell{4.82}
& \oodcell{0.60} & \oodcell{0.89}
& 97.46 & 98.84
& 84.83 & 89.55 \\

& Graph2SMILES+H~\citep{tu2022permutation}
& \oodcell{2.66} & \oodcell{4.32}
& \oodcell{0.54} & \oodcell{0.74}
& 96.52 & 97.64
& 82.37 & 87.59 \\

& MT~\citep{schwaller2019molecular}
& \oodcell{5.88} & \oodcell{9.39}
& \oodcell{0.30} & \oodcell{0.63}
& 92.73 & 94.90
& 74.38 & 81.34 \\

& FlowER~\citep{joung2025electron}
& \oodcell{18.86} & \oodcell{\underline{33.01}}
& \oodcell{2.41} & \oodcell{7.27}
& 96.99 & 99.02
& 84.09 & 92.74 \\

& DeepMech~\citep{das2026deepmech}
& \oodcell{\underline{24.93}} & \oodcell{32.13}
& \oodcell{\underline{5.60}} & \oodcell{\underline{11.56}}
& 98.98 & 99.58
& 95.94 & 96.98 \\

& \textbf{MechaVLM (ours)}
& \oodcell{\textbf{28.37}} & \oodcell{\textbf{46.25}}
& \oodcell{\textbf{6.76}} & \oodcell{\textbf{14.29}}
& 98.38 & 99.47
& 96.32 & 98.21 \\

\bottomrule
\end{tabular}
}
\vspace{-2 em}
\end{table*}

\begin{table*}[!htbp]
\centering
\footnotesize
\caption{
Zero-shot prediction on FukuyamaBench and MechBench (\%).
Models are trained on the source dataset and evaluated
without target-domain adaptation. $^*$: derived from~\citep{dang2026learning}
}
\label{tab:fukuyama_mechbench}
\setlength{\tabcolsep}{6pt}
\renewcommand{\arraystretch}{1.0}

\resizebox{\textwidth}{!}{
\begin{tabular}{@{}llcccccccc@{}}
\toprule
&
&
\multicolumn{4}{c}{\textbf{FukuyamaBench}}
&
\multicolumn{4}{c}{\textbf{MechBench}} \\
\cmidrule(lr){3-6}
\cmidrule(lr){7-10}

&
&
\multicolumn{2}{c}{Step}
&
\multicolumn{2}{c}{Pathway}
&
\multicolumn{2}{c}{Step}
&
\multicolumn{2}{c}{Pathway} \\
\cmidrule(lr){3-4}
\cmidrule(lr){5-6}
\cmidrule(lr){7-8}
\cmidrule(lr){9-10}

\textbf{Source}
& \textbf{Method}
& Top-1 & Top-5
& Top-1 & Top-5
& Top-1 & Top-5
& Top-1 & Top-5 \\
\midrule

\multirow{11}{*}{ReactMech}

& {\emph{Closed-book}} \\

& NERF~\citep{bi2021non}
& 6.56 & 8.86
& 0.31 & 0.31
& 0.72 & 1.20
& 0.14 & 0.18 \\

& Graph2SMILES~\citep{tu2022permutation}
& 2.35 & 4.01
& 0.31 & 0.31
& 0.14 & 0.36
& 0.00 & 0.05 \\

& Molecular Transformer~\citep{schwaller2019molecular}
& 3.41 & 7.51
& 0.00 & 0.63
& 0.39 & 1.07
& 0.09 & 0.14 \\

& FlowER~\citep{joung2025electron}
& 0.35 & 2.80
& 0.63 & 0.63
& \underline{4.62} & \underline{6.40}
& \underline{0.32} & \underline{0.32} \\

& DeepMech~\citep{das2026deepmech}
& \underline{9.01} & \underline{15.27}
& \underline{1.57} & \underline{1.57}
& 2.19 & 3.86
& 0.14 & \underline{0.32} \\

& \textbf{MechaVLM (ours)}
& \textbf{13.72} & \textbf{19.28}
& \textbf{3.13} & \textbf{3.13}
& \textbf{6.17} & \textbf{7.06}
& \textbf{1.01} & \textbf{1.37} \\

\cmidrule(lr){2-10}

& {\emph{Open-book}} \\

& ChemVLR-8B + CLIP retrieval
  ~\citep{radford2021learning}
& 10.32 & 14.97
& 1.57 & 1.88
& 3.52 & 4.89
& 0.46 & 0.64 \\

& ChemVLR-8B + OCSR retrieval
  ~\citep{fan2024openchemie}
& \underline{10.92} & \underline{15.47}
& \underline{1.88} & \underline{2.19}
& \underline{3.79} & \underline{5.18}
& \underline{0.60} & \underline{0.73} \\

& \textbf{MechaVLM (ours)}
& \textbf{16.83} & \textbf{22.93}
& \textbf{4.08} & \textbf{5.33}
& \textbf{8.48} & \textbf{9.96}
& \textbf{1.88} & \textbf{2.38} \\

\midrule

\multirow{16}{*}{FlowER}

& {\emph{Closed-book}} \\

& NERF~\citep{bi2021non}
& 14.07 & 18.88
& 1.25 & 1.25
& 3.21 & 4.93
& 0.23 & 0.27 \\

& Graph2SMILES~\citep{tu2022permutation}
& \underline{16.22} & 21.53
& 1.25 & 1.25
& 1.83 & 4.76
& 0.37 & 0.64 \\

& Molecular Transformer~\citep{schwaller2019molecular}
& 14.27 & 19.38
& 0.94 & 2.51
& 2.54 & 2.93
& 0.18 & 0.23 \\

& FlowER~\citep{joung2025electron}
& 13.92 & \underline{24.84}
& 1.57 & 2.51
& 6.18 & \underline{8.09}
& 0.14 & 0.23 \\

& Qwen3-235B-A22B$^*$~\citep{yang2025qwen3}
& 10.07 & --
& 4.08 & 6.90
& -- & --
& -- & -- \\

& Qwen3-30B-A3B$^*$~\citep{yang2025qwen3}
& 2.35 & --
& 3.45 & 4.39
& -- & --
& -- & -- \\

& ChemDFM-R-14B$^*$~\citep{zhao2025chemdfm}
& 0.70 & 0.70
& 3.13 & 3.76
& -- & --
& -- & -- \\

& ChemVLM-8B~\citep{li2025chemvlm}
& 8.01 & 11.57
& 0.00 & 0.00
& 4.12 & 5.64
& 0.18 & 0.41 \\

& TinyChemVL~\citep{zhao2026tinychemvl}
& 11.52 & 15.87
& 0.31 & 0.63
& 5.31 & 6.92
& 0.46 & 0.82 \\

& ChemVLR-8B~\citep{zhao2026chemvlr}
& 13.52 & 18.18
& 0.31 & 0.94
& \underline{6.36} & 7.89
& \underline{0.73} & \underline{1.24} \\

& \textbf{MechaVLM (ours)}
& \textbf{22.28} & \textbf{28.24}
& \textbf{5.96} & \textbf{7.52}
& \textbf{9.59} & \textbf{12.46}
& \textbf{1.56} & \textbf{1.92} \\

\cmidrule(lr){2-10}

& {\emph{Open-book}} \\

& ChemVLR-8B + CLIP retrieval
  ~\citep{radford2021learning}
& 14.72 & 19.33
& 0.63 & 1.25
& 6.88 & 8.43
& 0.82 & 1.37 \\

& ChemVLR-8B + OCSR retrieval
  ~\citep{fan2024openchemie}
& \underline{14.92} & \underline{20.08}
& \underline{0.94} & \underline{1.25}
& \underline{7.24} & \underline{8.60}
& \underline{1.01} & \underline{1.47} \\

& \textbf{MechaVLM (ours)}
& \textbf{27.14} & \textbf{33.40}
& \textbf{8.15} & \textbf{9.72}
& \textbf{12.72} & \textbf{15.41}
& \textbf{2.52} & \textbf{3.16} \\

\bottomrule
\end{tabular}
}
\vspace{-1 em}
\end{table*}

\subsection{Zero-Shot Transfer to Unseen Chemistry}
\label{sec:ood}

We evaluate whether mechanism predictors trained on one reaction distribution
transfer to unseen chemistry without target-domain adaptation, under both
closed- and open-book settings.

\textbf{Strong in-domain performance does not imply transferability.}
Tab.~\ref{tab:cross_domain} reveals a clear separation between ID accuracy and
cross-domain robustness.
With FlowER training, MechaVLM improves the strongest baseline by only
0.93/1.38  in Step/Pathway Top-1 on the FlowER test set, whereas the
margins increase to 12.50/13.93 after transfer to ReactMech.
The same trend holds in reverse: despite lower ID Step accuracy on ReactMech, MechaVLM achieves the strongest zero-shot transfer to FlowER.
Meanwhile, chemistry VLMs that are competitive in-domain deteriorate much more
sharply under distribution shift.
These results suggest that the advantage of MechaVLM primarily comes from
better transfer of mechanistic knowledge rather than improved source-domain
fitting.

\begin{figure}[t]
    \centering
    \includegraphics[width=1.0\textwidth]{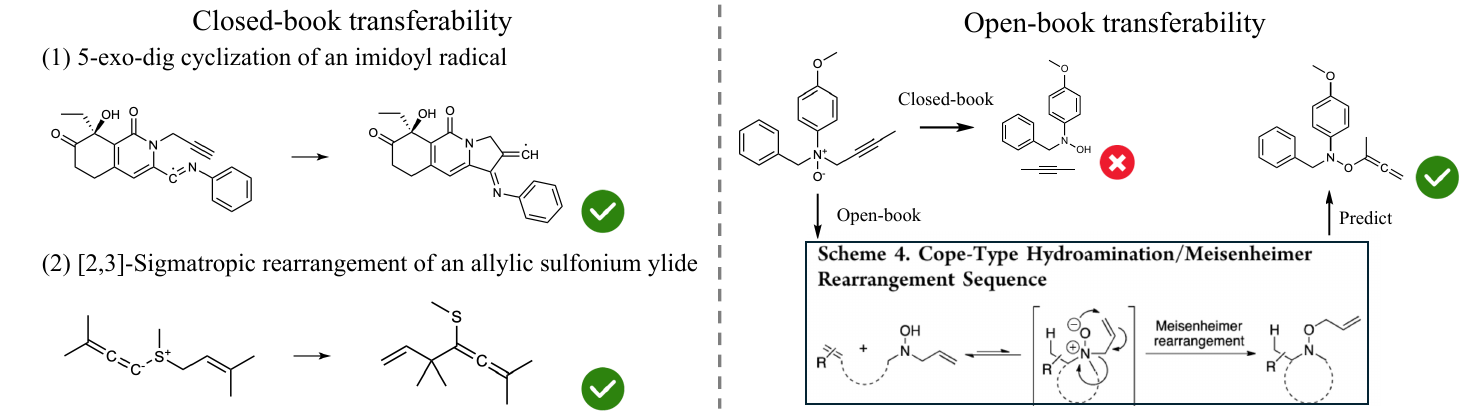}
    \vspace{-2em}
    \caption{Examples of closed-book and open-book prediction of MechaVLM for OOD reactions.
    }
    \label{fig:examples}
    \vspace{-1.5em}
\end{figure}

\textbf{Held-out benchmarks expose a more severe generalization gap.}
We further evaluate on held-out FukuyamaBench and MechBench.
As in Tab.~\ref{tab:fukuyama_mechbench}, performance decreases
substantially for all methods, with the drop being particularly pronounced for
complete pathway prediction.
This indicates that transferring individual elementary transformations is
already challenging, while errors further accumulate during multi-step
mechanism rollout.
Despite this harder setting, MechaVLM consistently achieves the strongest
closed-book transfer. Fig.~\ref{fig:examples} (left) shows two examples.

\textbf{External precedents provide complementary knowledge under distribution shift.}
Open-book prediction further improves both held-out benchmarks. Fig.~\ref{fig:examples} (right) shows an example.
With FlowER training, Step/Pathway Top-1 increases from 22.28/5.96 to
27.14/8.15 on FukuyamaBench and from 9.59/1.56 to 12.72/2.52 on MechBench,
with similar gains from ReactMech.
MechaVLM also outperforms chemistry VLMs with visual or
OCSR-based retrieval, indicating that external precedents are most effective
when their mechanistic evidence is transferred to the query
(Sec.~\ref{sec:ablation}).

\begin{table*}[t]
\centering
\scriptsize
\renewcommand{\arraystretch}{1.0}

\begin{minipage}[t]{0.47\textwidth}
    \centering
    \caption{Transfer to atom mapping (\%). }
    \label{tab:pretrain_ablation}
    \setlength{\tabcolsep}{1pt}

    \begin{tabular}{@{}lccc@{}}
    \toprule[1.5pt]
    \textbf{Method}
    & \shortstack{\textbf{USPTO}\\\textbf{-50K (2K)}}
    & \textbf{Schneider}
    & \textbf{Jaworski} \\
    \midrule[1pt]

    \multicolumn{4}{l}{\emph{Specialized atom-mapping methods}} \\
    Indigo~\citep{indigo_toolkit_2026}
        & 30.35 & 38.98 & 15.93 \\
    G.Mapper~\citep{nugmanov2022bidirectional}
        & 79.40 & 92.20 & 80.40 \\
    RxnMapper~\citep{schwaller2021extraction}
        & \underline{81.05} & \underline{93.12} & 83.15 \\
    LocalMapper~\citep{chen2024precise}
        & -- & 90.08 & \underline{86.98} \\

    \midrule
    \multicolumn{4}{l}{\emph{Vision-language models}} \\
    Qwen2.5-VL-3B (Base VLM)
        & 72.23 & 84.36 & 79.63 \\
    ChemVLM-8B~\citep{li2025chemvlm}
        & 72.94 & 84.91 & 80.21 \\
    TinyChemVL~\citep{zhao2026tinychemvl}
        & 74.38 & 85.76 & 81.47 \\
    ChemVLR-8B~\citep{zhao2026chemvlr}
        & 75.86 & 86.63 & 82.35 \\

    \midrule
    \textbf{MechaVLM(ours)}
        & \textbf{83.62}
        & \textbf{93.14}
        & \textbf{89.72} \\

    \bottomrule[1.5pt]
    \end{tabular}
\end{minipage}
\begin{minipage}[t]{0.51\textwidth}
    \centering
    \caption{Transfer to reaction center prediction.}
    \label{tab:reaction_center_ablation}
    \setlength{\tabcolsep}{1pt}

    \begin{tabular}{@{}lcccc@{}}
    \toprule[1.5pt]
    &
    \multicolumn{2}{c}{\textbf{Class Unknown}}
    &
    \multicolumn{2}{c}{\textbf{Class Known}} \\
    \cmidrule(lr){2-3}
    \cmidrule(lr){4-5}

    \textbf{Method}
    & Top-1 & Top-5
    & Top-1 & Top-5 \\
    \midrule[1pt]

    \multicolumn{5}{l}{\emph{Specialized reaction-center methods}} \\
    RetroXpert~\citep{yan2020retroxpert}
        & 64.9 & -- & 86.0 & -- \\
    G2Gs~\citep{shi2020graph}
        & 75.8 & 85.6 & 90.2 & 95.0 \\
    GDiffRetro~\citep{sun2025gdiffretro}
        & 86.2 & \underline{98.8} & -- & -- \\

    \midrule
    \multicolumn{5}{l}{\emph{Vision-language models}} \\
    Qwen2.5-VL-3B (Base VLM)
        & 89.6 & 93.1 & 92.7 & 94.3 \\
    ChemVLM-8B~\citep{li2025chemvlm}
        & 90.8 & 94.2 & 93.8 & 95.4 \\
    TinyChemVL~\citep{zhao2026tinychemvl}
        & 92.1 & 95.7 & 95.0 & 96.6 \\
    ChemVLR-8B~\citep{zhao2026chemvlr}
        & \underline{94.0}
        & 97.4
        & \underline{96.1}
        & \underline{98.0} \\

    \midrule
    \textbf{MechaVLM(ours)}
        & \textbf{97.2}
        & \textbf{99.3}
        & \textbf{98.4}
        & \textbf{99.6} \\

    \bottomrule[1.5pt]
    \end{tabular}
\end{minipage}

\vspace{-2.2em}
\end{table*}

\begin{table*}[t]
\centering
\footnotesize
\caption{
Cumulative ablation of MechaVLM (Top-1 \%).
All variants are trained on FlowER.
White cells report in-distribution performance, while gray cells denote
zero-shot OOD evaluation.
}
\label{tab:main_ablation}
\setlength{\tabcolsep}{9pt}
\renewcommand{\arraystretch}{1.}

\resizebox{\textwidth}{!}{
\begin{tabular}{@{}lcccccccc@{}}
\toprule
&
\multicolumn{2}{c}{\textbf{FlowER (ID)}}
&
\multicolumn{2}{c}{\textbf{ReactMech}}
&
\multicolumn{2}{c}{\textbf{FukuyamaBench}}
&
\multicolumn{2}{c}{\textbf{MechBench}} \\
\cmidrule(lr){2-3}
\cmidrule(lr){4-5}
\cmidrule(lr){6-7}
\cmidrule(lr){8-9}

\textbf{Variant}
& Step & Path.
& Step & Path.
& Step & Path.
& Step & Path. \\
\midrule

\multicolumn{9}{l}{\emph{Input modality}} \\

SMILES input
& 87.85 & 90.92
& \oodcell{43.38} & \oodcell{23.88}
& \oodcell{10.67} & \oodcell{0.63}
& \oodcell{4.31} & \oodcell{0.18} \\

Graph input
& 88.35 & 91.36
& \oodcell{45.87} & \oodcell{25.91}
& \oodcell{12.32} & \oodcell{0.94}
& \oodcell{5.08} & \oodcell{0.32} \\
\midrule

\multicolumn{9}{l}{\emph{Closed-book: transferable representation learning}} \\

Mechanism SFT only (image input)
& 88.41 & 91.96
& \oodcell{52.32} & \oodcell{32.47}
& \oodcell{15.87} & \oodcell{1.57}
& \oodcell{6.82} & \oodcell{0.60} \\

+ Multiscale chemical grounding
& 89.63 & 93.42
& \oodcell{62.74} & \oodcell{45.58}
& \oodcell{19.58} & \oodcell{3.76}
& \oodcell{8.58} & \oodcell{1.10} \\

+ Cross-rendering contrastive learning
& 90.02 & 93.89
& \oodcell{69.34} & \oodcell{53.23}
& \oodcell{22.28} & \oodcell{5.96}
& \oodcell{9.59} & \oodcell{1.56} \\

\midrule

\multicolumn{9}{l}{\emph{Open-book: visual precedent utilization}} \\

+ Retrieved precedent context
& 89.91 & 93.78
& \oodcell{70.61} & \oodcell{54.72}
& \oodcell{23.59} & \oodcell{6.27}
& \oodcell{10.08} & \oodcell{1.60} \\

+ Initial-state cross-image transfer
& 90.08 & \textbf{93.96}
& \oodcell{72.54} & \oodcell{56.83}
& \oodcell{24.89} & \oodcell{6.90}
& \oodcell{10.54} & \oodcell{1.69} \\

+ Step-wise cross-image transfer
& 89.97 & 93.85
& \oodcell{75.09} & \oodcell{60.18}
& \oodcell{26.39} & \oodcell{7.52}
& \oodcell{11.96} & \oodcell{2.15} \\

\textbf{+ Token-wise gated fusion (MechaVLM)}
& \textbf{90.11} & 93.88
& \oodcell{\textbf{76.21}} & \oodcell{\textbf{61.52}}
& \oodcell{\textbf{27.14}} & \oodcell{\textbf{8.15}}
& \oodcell{\textbf{12.72}} & \oodcell{\textbf{2.52}} \\

\bottomrule
\end{tabular}
}
\vspace{-1.5em}
\end{table*}

\vspace{-0.5em}
\subsection{Transferable Chemical Representations Beyond Mechanism Prediction}
\label{sec:downstream_transfer}
\vspace{-0.5em}

We further explore whether the knowledge learned by MechaVLM transfers beyond
mechanism prediction to broader reaction-related tasks: atom mapping and reaction
center prediction. Molecular inputs are rendered into 2D images (see Appendix) for training and test in VLM-based models.

\textbf{Atom mapping.}
Following~\citep{chen2024precise}, we adapt VLM-based models on USPTO-FULL
and evaluate on a held-out 2K subset of USPTO-50K, Schneider, and Jaworski.
As in Tab.~\ref{tab:pretrain_ablation}, MechaVLM outperforms the base VLM without MechaVLM training, existing chemistry VLMs, and matches or surpasses specialized atom-mapping methods. These may be attributed to a twofold advantage: the pre-training phase improves the model’s visual comprehension of 2D molecular geometries, and simultaneously transfers a powerful mechanistic prior for complex atom mapping.

\textbf{Reaction center prediction.}
Tab.~\ref{tab:reaction_center_ablation} further validates this dual benefit for identifying reaction sites on USPTO-50K. The test set is divided with no overlap with training set of MechaVLM. Pre-training significantly improves MechaVLM’s Top-1 accuracy, outperforming strong graph-based baselines like G2Gs. This may be because our pre-training explicitly conditions the model to pinpoint active sites prior to predicting electron flows, intrinsically optimizing the model’s localized visual attention.

\subsection{Ablation and Analysis}
\label{sec:ablation}
\vspace{-0.5em}

We ablate the main components of MechaVLM with FlowER as the training source.
We report Top-1 Step/Pathway accuracy on FlowER and three OOD
benchmarks. More analyses are in Appendix.

\textbf{Visual inputs improve cross-domain transfer under matched training.}
As in Tab.~\ref{tab:main_ablation}, with the same mechanism supervision, SMILES, graph, and image inputs perform
similarly on FlowER but diverge substantially under OOD evaluation.
Image input achieves the strongest transfer across ReactMech, FukuyamaBench,
and MechBench, indicating that visual representations provide a better basis
for cross-domain generalization even before additional representation learning
or retrieval.

\textbf{Representation learning drives closed-book transfer.}
As in Tab.~\ref{tab:main_ablation}, adding multiscale chemical grounding
and cross-rendering contrastive learning improves FlowER Step/Pathway accuracy
by only 1.61/1.93 points, but yields 17.02/20.76-point gains after transfer to
ReactMech, with consistent improvements on both held-out benchmarks.
The larger gains under distribution shift indicate that the learned features remain useful beyond the source reaction distribution.
This suggests that these objectives mainly improve the transferability of the
learned representation rather than source fitting.

\textbf{Step-wise transfer makes better use of visual precedents.}
Using the same retrieved precedents, direct context addition gives only modest
gains, while cross-image transfer performs better by matching the current state
to relevant precedent regions.
Recomputing this transfer at each mechanism step improves further, and
token-wise gated fusion gives the best results.
This supports retrieving precedents once but updating the used evidence as the
mechanism evolves.

\textbf{Coarse-to-fine retrieval improves precedent selection.}
Tab.~\ref{tab:retriever_ablation} shows that our learned visual representations
outperform both CLIP and OCSR-based retrieval.
Late interaction further improves over global similarity by capturing local
matches between the query reactants and regions of the full mechanism figure.
Combining global retrieval with late-interaction reranking performs best,
showing that overall reaction similarity and local structural matching provide
complementary signals.

\begin{table*}[t]
\centering
\footnotesize
\caption{
Ablation of external precedent retrieval strategies (\%).
All models are trained on FlowER.
}
\label{tab:retriever_ablation}
\setlength{\tabcolsep}{10pt}
\renewcommand{\arraystretch}{1.}

\resizebox{0.99\textwidth}{!}{
\begin{tabular}{@{}lcccccc@{}}
\toprule
&
\multicolumn{2}{c}{\textbf{ReactMech}}
&
\multicolumn{2}{c}{\textbf{FukuyamaBench}}
&
\multicolumn{2}{c}{\textbf{MechBench}} \\
\cmidrule(lr){2-3}
\cmidrule(lr){4-5}
\cmidrule(lr){6-7}

\textbf{Retriever}
& Step & Path.
& Step & Path.
& Step & Path. \\
\midrule

Random
& 68.52 & 52.41
& 21.83 & 5.64
& 9.38 & 1.47 \\

CLIP image similarity~\citep{radford2021learning}
& 70.74 & 54.63
& 23.44 & 6.27
& 10.17 & 1.60 \\

OCSR structure similarity~\citep{fan2024openchemie}
& 71.08 & 55.12
& 23.74 & 5.96
& 10.37 & 1.65 \\

\midrule

Global similarity
& 72.93 & 56.76
& 24.89 & 6.90
& 11.31 & \underline{2.15} \\

Late interaction
& \underline{74.41} & \underline{58.92}
& \underline{25.99} & \underline{7.52}
& \underline{12.13} & \underline{2.15} \\

\textbf{Global $\rightarrow$ Late Interaction (ours)}
& \textbf{76.21} & \textbf{61.52}
& \textbf{27.14} & \textbf{8.15}
& \textbf{12.72} & \textbf{2.52} \\

\bottomrule
\end{tabular}
}
\vspace{-1em}
\end{table*}

\vspace{-0.5em}
\section{Conclusion}
\vspace{-0.5em}

We introduced \textbf{MechaVLM}, a visual framework for zero-shot reaction
mechanism transfer across chemical domains.
MechaVLM combines transferable visual representations with external visual
precedents, achieving state-of-the-art zero-shot performance across
cross-dataset and literature-derived benchmarks.
Its learned representation also transfers to atom mapping and reaction center
prediction, while external precedents provide further gains in the open-book
setting.
A remaining limitation is the dependence on the coverage and quality of the
external precedent bank, motivating future work on broader and more reliable
mechanistic knowledge sources.

\subsection*{AI use statement}

In this work, we used generative AI tools to provide feedback on research methodology and experimental design, assist with the interpretation and presentation of experimental results, and support parts of the data annotation process. In particular, generative AI was used to assist the extraction and transcription of reaction-mechanism information from literature figures, after which all annotations were manually reviewed and verified by the authors or domain experts. We did not use generative AI tools to generate synthetic datasets, implement the proposed methods, formulate or prove mathematical claims, or perform experiments. We also used generative AI tools to assist with manuscript drafting, language polishing, structural organization, and readability. All AI-assisted suggestions and annotations were critically reviewed, and all experimental results and scientific claims were verified against the underlying data and outputs. We take responsibility for the final content of this work, including text, claims, annotations, and other artifacts produced with the aid of generative AI.

\subsection*{Ethics statement}

This work uses existing chemical datasets and published scientific literature and does not involve human subjects, personal data, or user studies. MechBench and the external precedent bank are derived from published literature; any release will respect applicable copyright and licensing restrictions and preserve source provenance. We also remove overlaps across training, retrieval, and evaluation data to reduce leakage. Reaction mechanism prediction may benefit chemical understanding, synthesis planning, and scientific discovery, but may also have dual-use applications involving hazardous chemistry. This work is intended for research and education, and we encourage responsible use under relevant legal, safety, and ethical standards.

\subsection*{Reproducibility statement}

We provide detailed descriptions of the model architecture, training procedure,
datasets, evaluation protocols, and implementation settings in the main text
and Appendix. The Appendix further documents the construction and processing of
MechBench and the external precedent bank, together with additional experimental
details and ablations. Code and benchmark resources will be released to support
reproduction of the reported results. For the external precedent bank, we will
provide the source DOI, figure index, cropping coordinates, and reconstruction
scripts for each retained figure, enabling the corpus to be rebuilt from the
original publications.

\bibliography{iclr2027_conference}
\bibliographystyle{iclr2027_conference}

\appendix

\section{MechBench: A Literature-Derived Mechanism Benchmark}
\label{app:mechbench}

\textbf{Motivation.}
Existing reaction-mechanism benchmarks are primarily derived from curated
databases or educational sources.
While these datasets provide well-controlled evaluation settings, they cover
only a limited subset of the mechanisms encountered in the broader chemical
literature.
To evaluate zero-shot transfer under a more realistic distribution shift, we
construct \textbf{MechBench}, a literature-derived benchmark collected from
published chemistry articles.
MechBench contains \textbf{2,184 complete mechanism pathways} comprising
\textbf{9,146 elementary steps}, collected from \textbf{417 journal publications}.
The benchmark is used exclusively for evaluation and is never used for model
training or target-domain adaptation.

\textbf{Literature collection.}
We manually collected publications from three chemistry journals:
\emph{Journal of the American Chemical Society} (JACS),
\emph{The Journal of Organic Chemistry} (JOC), and
\emph{Organic Letters} (OL).
Within the 2000--2019 publication period, we first searched for articles whose
titles contained the terms \emph{synthesis} or \emph{reaction}, and then
manually screened the retrieved papers for explicit reaction-mechanism figures.
We retained only figures from the main article that contained a complete,
self-contained mechanistic pathway, such that the intermediates and elementary
transformations required for pathway reconstruction were available directly
from the figure without relying on Supporting Information.
This procedure yielded mechanisms from 417 journal publications.

\textbf{Mechanism curation and quality control.}
For each retained figure, GPT-6-Astra~\citep{openai2026api} was used to assist the extraction and
transcription of the depicted mechanism into an ordered sequence of molecular
intermediates and elementary transformations.
Atom mappings between consecutive states were assigned using
RxnMapper~\citep{schwaller2021extraction}, from which the corresponding
bond-order and atom-local electron changes were derived.
The resulting pathways were subsequently checked against the original figures by three chemical experts until agreement.
We retained only examples for which a complete and evaluable sequence of
intermediates could be reconstructed, excluding incomplete, duplicated, or
ambiguous pathways.

\textbf{Held-out evaluation and leakage prevention.}
MechBench is treated as a fully held-out benchmark.
No MechBench examples are used for representation learning, mechanism
supervised fine-tuning, or target-domain adaptation.
For the open-book setting, benchmark reactions are additionally excluded from
the external precedent corpus, preventing retrieval of the evaluated mechanism
itself.
Thus, both closed-book and open-book results measure transfer to unseen
mechanism instances rather than memorization of benchmark examples.

\begin{figure}[t]
    \centering
    \includegraphics[width=1.0\textwidth]{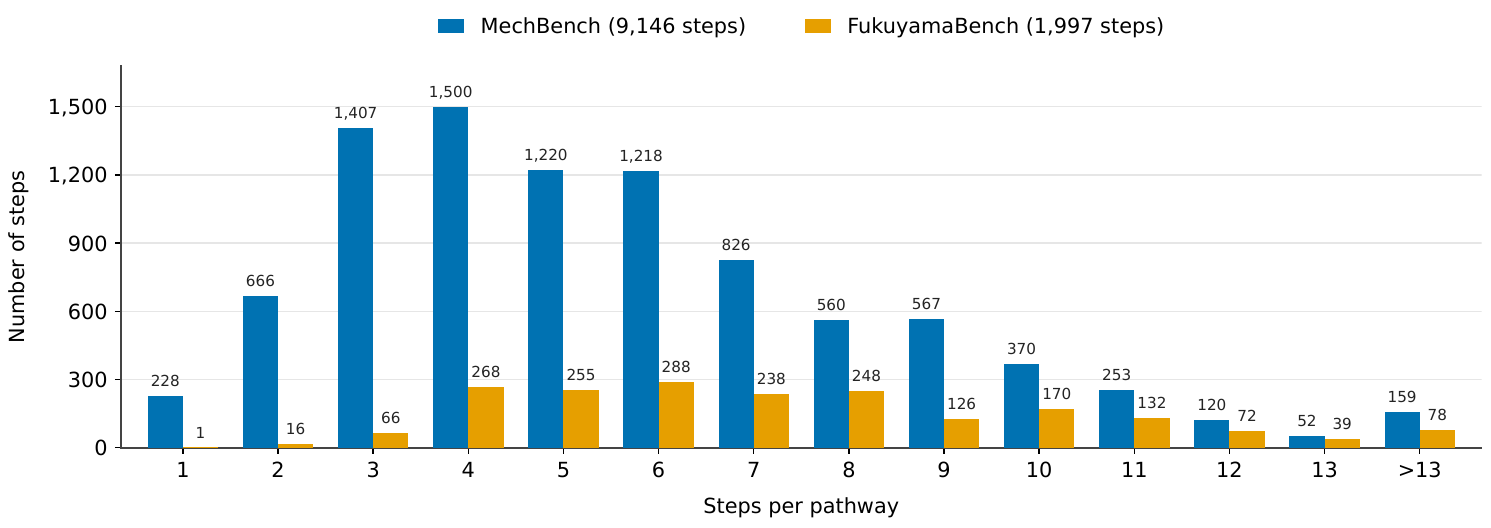}
    \vspace{-1em}
    \caption{
    Distribution of elementary steps by pathway length in MechBench and FukuyamaBench. Bars indicate the
  total number of elementary steps contributed by pathways of each length.
    }
    \label{fig:mechbench}
\end{figure}

\textbf{Comparison with FukuyamaBench.}
As shown in Fig.~\ref{fig:mechbench} and Tab.~\ref{tab:benchmark_statistics}, MechBench complements FukuyamaBench, which contains 319 expert-curated
mechanism pathways and 1,997 elementary steps from the Fukuyama Mechanism
Book.
Whereas FukuyamaBench provides a compact textbook-derived evaluation,
MechBench draws mechanisms from hundreds of journal publications and contains
approximately $6.85\times$ as many pathways and $4.58\times$ as many elementary
steps.
The two benchmarks therefore provide complementary tests of zero-shot
mechanism transfer beyond the source training distribution.

\begin{table}[t]
\centering
\small
\caption{
Comparison of the fully held-out mechanism benchmarks used in this work.
Neither benchmark is used for training or target-domain adaptation.
}
\label{tab:benchmark_statistics}
\setlength{\tabcolsep}{6pt}
\renewcommand{\arraystretch}{1.08}
\begin{tabular}{@{}lcccc@{}}
\toprule
\textbf{Benchmark} &
\textbf{Source} &
\textbf{\# Publications} &
\textbf{\# Pathways} &
\textbf{\# Steps} \\
\midrule
FukuyamaBench &
Mechanism book &
-- &
319 &
1,997 \\
MechBench &
Journal literature &
417 &
2,184 &
9,146 \\
\bottomrule
\end{tabular}
\end{table}

\section{External Visual Precedent Bank}
\label{app:precedent_bank}

\textbf{Literature collection and figure screening.}
We construct a fixed external visual precedent bank from five publication
sources: \emph{Journal of the American Chemical Society} (JACS),
\emph{The Journal of Organic Chemistry} (JOC),
\emph{Organic Letters} (OL), \emph{Nature Communications}, and ChemRxiv.
We first identify publications whose titles contain the terms
\emph{synthesis} or \emph{reaction}.
The publications are rendered as page images, from which candidate figures are
cropped using VisualHeist~\citep{leong2025mermaid}.
We then use Qwen2.5-VL-32B~\citep{bai2025qwen25vltechnicalreport} to screen the
candidate crops with explicit inclusion criteria.
A figure is retained only when it depicts a complete and self-contained
reaction mechanism, such that the molecular structures, relevant intermediates,
and mechanistic transformations are available
within the figure.
The final bank contains \textbf{70,384 mechanism figures}.

\textbf{Native visual representation.}
All selected figures are retained as their original figure crops.
We do not convert them to SMILES, molecular graphs, or other symbolic
representations, nor do we explicitly parse intermediates, electron-flow arrows,
or textual annotations.
Consequently, molecular structures, curved arrows, reaction annotations, and
other visual information remain available to the model in their original form.
The precedent bank provides no structured mechanism labels or
query--precedent correspondence annotations and is used purely as external
visual memory.

\textbf{Deduplication and leakage prevention.}
We first remove visually near-duplicate figure crops using
CLIP~\citep{radford2021learning} similarity, discarding one item from pairs with
cosine similarity above $0.95$.
We intentionally retain mechanistically related but distinct reactions, since
such recurring precedents constitute useful external chemical knowledge.
To prevent evaluation leakage, all publications used to construct MechBench are
excluded from the precedent bank at the DOI level.
We additionally perform reaction-instance overlap screening against the
supervised mechanism datasets and held-out benchmarks.
Qwen2.5-VL-32B is used to flag candidate overlaps based on the depicted
molecular structures and reactions, and flagged cases are manually inspected
before inclusion.
This procedure removes identical reaction instances while retaining related
mechanistic patterns that differ in their molecular realization.

\textbf{Fixed external memory.}
The precedent bank is constructed before mechanism training and remains fixed
thereafter.
The same bank is accessible during source-domain mechanism training and
target-domain open-book inference, but its figures are never used as supervised
mechanism targets.
Thus, the model can learn how to use retrieved external evidence during
training without receiving target-domain examples or updating the external
memory at test time.
In the closed-book setting, the precedent bank is disabled entirely.

\textbf{Offline indexing and retrieval.}
After Stage-I representation learning, the frozen visual encoder
$\mathcal{E}_{\theta_0}$ is used to pre-compute both the global descriptor
$z(I_i^p)$ and dense visual features $H_i^p$ for every precedent.
These representations are cached and reused throughout training and inference,
so only the current query image needs to be encoded online.
Following Sec.~\ref{sec:retrieval}, retrieval is performed in two stages.
We first use global descriptor similarity to retrieve the top
$L=64$ candidates and then rerank them using dense late interaction, retaining
the top $K=4$ precedents for cross-image transfer.
Unless otherwise specified, we use these values throughout the experiments.

\begin{figure}[t]
    \centering
    \includegraphics[width=1.0\textwidth]{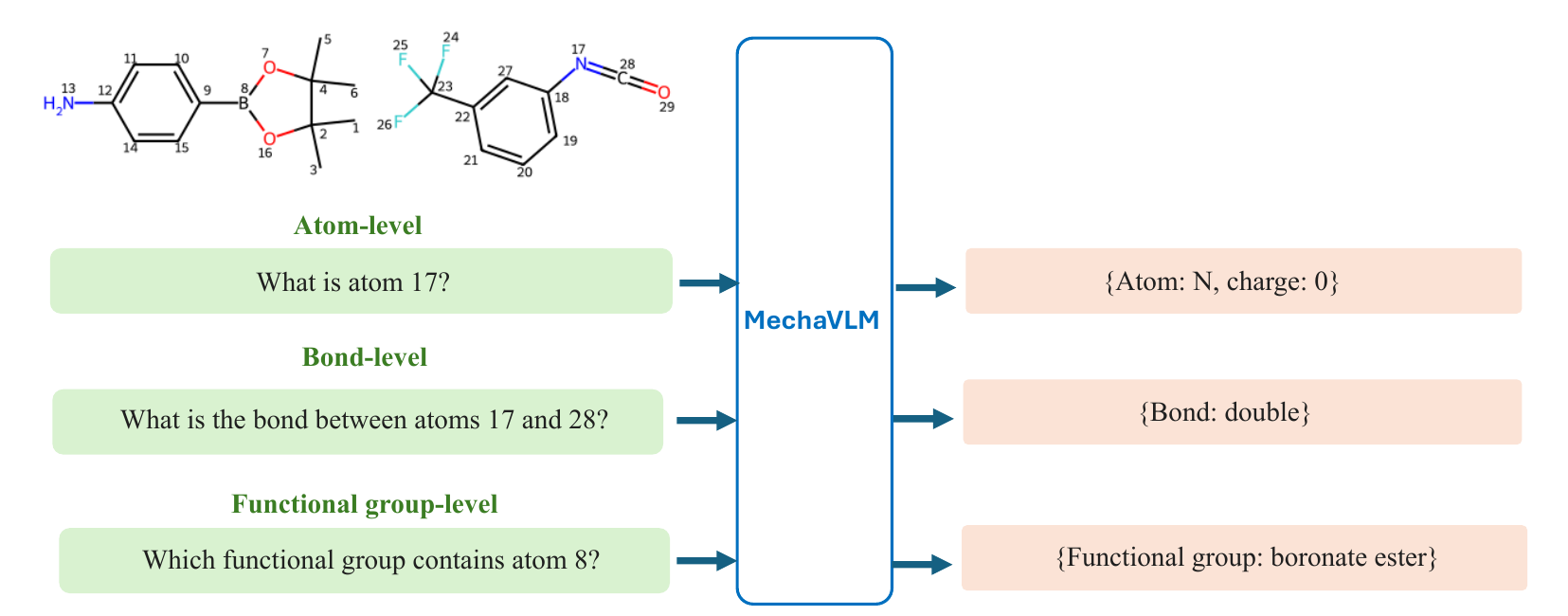}
    \vspace{-2em}
    \caption{
    Example of multiscale chemical grounding.
    }
    \label{fig:grounding}
\end{figure}

\section{Multiscale Chemical Grounding}
\label{app:chemical_grounding}

The multiscale grounding objective in Sec.~\ref{sec:visual_grounding} is
generated automatically from the structured reactants of the corresponding
source-domain training set.
No additional molecular corpus, product information, reaction-center labels, or
mechanism annotations are used.
The objective grounds visual representations at three complementary scales:
individual atoms, local connectivity, and functional-group environments.

\textbf{Visual indexing.}
All rendered heavy atoms across all reactant molecules are assigned globally
unique visual indices.
The indices serve only as visual anchors for referring to specific chemical
units and are independently randomized for every rendering.
Original atom-mapping identifiers are never exposed to the model, preventing
persistent indices from becoming shortcuts across examples or rendering
variants.
Implicit hydrogens are not assigned indices.

\textbf{Grounding tasks.}
Given an indexed reactant image, we construct the three instruction-based tasks
summarized in Fig.~\ref{fig:grounding}.
Atom grounding queries a single indexed atom and requires its elemental identity
and formal charge.
Bond grounding queries a pair of indexed atoms and predicts their connectivity
and bond type, with bonded and non-bonded pairs sampled at approximately equal
frequency.
Functional-group grounding also queries a single indexed atom, but asks the
model to identify a functional group containing that atom.
Functional-group labels are obtained automatically through SMARTS matching.
Because chemical motifs may overlap, an atom can have multiple valid
functional-group labels. All matched labels are treated as correct.

\textbf{Functional-group vocabulary.}
Functional-group supervision is generated using a fixed SMARTS-based vocabulary
covering common local chemical environments in organic reactions.
For each queried atom, we collect all vocabulary entries whose SMARTS matches
contain that atom.
When multiple patterns overlap, all corresponding labels are retained as valid
answers.
During training, one valid label is randomly sampled as the textual target for
each grounding instance. During evaluation, a prediction is considered correct
if it matches any valid label.
Atoms that do not participate in any predefined functional group are assigned
\texttt{none}.
The vocabulary is summarized in Tab.~\ref{tab:functional_group_vocab}.

\begin{table}[!htbp]
\centering
\small
\caption{
Functional-group vocabulary used for atom-anchored functional-group grounding.
Multiple entries may be valid for the same queried atom when SMARTS patterns
overlap.
}
\label{tab:functional_group_vocab}
\setlength{\tabcolsep}{5pt}
\renewcommand{\arraystretch}{1.08}
\begin{tabular}{@{}llp{8.2cm}@{}}
\toprule
\textbf{Category} & \textbf{\#} & \textbf{Canonical vocabulary} \\
\midrule

Unsaturated / aromatic
& 4
& alkene, alkyne, conjugated diene, aromatic ring \\

Oxygen-containing
& 8
& alcohol, phenol, ether, epoxide, peroxide, acetal, hemiacetal,
silyl ether \\

Carbonyl / acyl
& 15
& aldehyde, ketone, carboxylic acid, ester, lactone, amide, lactam,
imide, anhydride, acyl fluoride, acyl chloride, acyl bromide,
carbonate, carbamate, urea \\

Nitrogen-containing
& 13
& primary amine, secondary amine, tertiary amine, quaternary ammonium,
imine, enamine, oxime, nitrile, nitro, azo, diazonium, azide,
isocyanate \\

Sulfur-containing
& 9
& thiol, thioether, disulfide, sulfoxide, sulfone, sulfonamide,
sulfonate ester, thioester, sulfonyl halide \\

Boron-containing
& 3
& boronic acid, boronate ester, organotrifluoroborate \\

Halogen-containing
& 12
& alkyl fluoride, alkyl chloride, alkyl bromide, alkyl iodide,
vinyl fluoride, vinyl chloride, vinyl bromide, vinyl iodide,
aryl fluoride, aryl chloride, aryl bromide, aryl iodide \\

Phosphorus-containing
& 4
& phosphine, phosphine oxide, phosphonium, phosphate ester \\

\midrule
\textbf{Total}
& \textbf{68}
& \\

\bottomrule
\end{tabular}
\end{table}

\textbf{Training supervision.}
All grounding tasks use the standard text output of the VLM.
Grounding instances are sampled from chemical units naturally occurring in each
reactant image, with the number of queries per rendering capped to prevent large
molecules from dominating training.
For functional-group grounding, one valid matched label is randomly selected as
the supervision target whenever multiple labels apply to the queried atom.
The same underlying chemical unit can therefore appear with different visual
indices, layouts, rendering styles, and valid functional-group descriptions
across epochs.

\section{Training and Evaluation Details}
\label{app:training_evaluation}

\textbf{Mechanism training.}
Closed-book and open-book models are initialized from the same Stage-I checkpoint and independently trained on the same source-domain mechanism data for 2 epochs. The visual encoder remains frozen, while the language backbone and output head are fully fine-tuned; the open-book model additionally trains randomly initialized cross-image transfer and gated-fusion modules with access to the fixed precedent bank. We use AdamW with a learning rate of $3\times10^{-6}$ and cosine decay, with identical batch size and optimization budget for both settings.

\textbf{State rendering and rollout.}
During training, we use ground-truth state conditioning: the ground-truth
molecular state $S_q^{t-1,*}$ is rendered to construct the visual input for
step $t$, and the corresponding electron edit $\Delta_q^{t,*}$ is used as the
supervision target.
During inference, the predicted edit is instead applied recursively,
\begin{equation}
    S_q^t
    =
    S_q^{t-1}
    \oplus
    \widehat{\Delta}_q^t,
    \qquad
    I_q^{t+1}
    =
    \rho(S_q^t).
\end{equation}
Atom indices are kept consistent throughout a mechanism rollout, and a fixed
rendering style is used at inference time.
In the open-book setting, reaction-level precedents are retrieved only once
from the initial reactants; subsequent states re-read the same precedents
through the step-conditioned local-transfer mechanism in
Sec.~\ref{sec:retrieval}.

\textbf{Decoding.}
We use deterministic beam search with a beam size of 5.
At each mechanism step, candidate structured electron edits are decoded and
propagated through their corresponding molecular states.
Decoding terminates when the model produces the terminal no-edit output,
indicating that no further electron transformation is predicted.
The retained beams are used to compute Top-1 and Top-5 results without
stochastic sampling.

\textbf{Evaluation protocol.}
We report Step and Pathway Top-$k$ accuracy using a common evaluation pipeline
across all retrained methods.
FlowER and ReactMech models use their official data splits, while FukuyamaBench
and MechBench are used exclusively for zero-shot evaluation and never for
hyperparameter tuning or target-domain adaptation.
All trainable baselines are retrained from their official implementations on
the same source-domain splits.
Vision-language baselines receive the same reactant-only image inputs and are
adapted using the same mechanism-supervision protocol as MechaVLM.
For open-book comparisons, all retrieval-based methods access the same external
precedent bank and retrieve the same number of final precedents.

\textbf{Metrics.}
Following~\citep{joung2025electron}, we report \emph{Top-$k$ step} and \emph{Top-$k$ pathway accuracy}, with $k\in\{1,5\}$. Because the two metrics use different units of averaging, namely individual steps and complete pathways, pathway accuracy is not necessarily bounded by step accuracy. In particular, a small number of long and difficult pathways can contribute many incorrect steps and lower the step-level average, while each complete pathway contributes only one instance to pathway accuracy.

\section{Implementation Details}
\label{app:implementation}

\textbf{Backbone and visual features.}
We implement MechaVLM on Qwen2.5-VL-3B
~\citep{bai2025qwen25vltechnicalreport}.
We treat its visual tower together with the multimodal projector as the visual
encoder $\mathcal{E}_{\theta}$.
The dense representation $H$ corresponds to the projected visual tokens
immediately before they are consumed by the language backbone.
This representation is shared across chemical grounding, precedent retrieval,
and cross-image transfer.
For reaction-level retrieval, we obtain the global descriptor by mean-pooling
the dense visual tokens followed by $\ell_2$ normalization,
\begin{equation}
    z(I)
    =
    \operatorname{Norm}
    \left(
        \frac{1}{N_I}
        \sum_{j=1}^{N_I} h_j
    \right).
\end{equation}
After Stage-I representation pretraining, the complete visual encoder
$\mathcal{E}_{\theta_0}$ is frozen.

\textbf{Precedent retrieval.}
Because the external precedent bank is fixed, we pre-compute and cache both
the normalized global descriptor $z(I_i^p)$ and dense visual features $H_i^p$
for every precedent using $\mathcal{E}_{\theta_0}$.
At inference, only the current query image is encoded online.
For the coarse stage, we compute exact cosine similarities against all cached
global descriptors and retain the top $L=64$ candidates.
We then perform exact late-interaction reranking using the $\ell_2$-normalized
dense visual features, following Sec.~\ref{sec:retrieval}, and retain the top
$K=4$ precedents for mechanism transfer.
The same retrieval configuration is used during mechanism training and
open-book evaluation.

\textbf{Cross-image transfer and fusion.}
For open-book prediction, the dense visual tokens of the retrieved precedents
are concatenated and accessed through a single multi-head cross-attention
module.
We use the current query features as queries and the precedent features as
keys and values.
The attention module uses 8 heads, with its hidden dimension matched to that of
the projected visual features.
The transferred representation is projected back to the query feature space
and incorporated using the token-wise gate defined in
Sec.~\ref{sec:retrieval}.
The cross-attention, output projection, and gating layers are optimized during open-book mechanism training, while the
visual encoder and cached precedent representations remain fixed.

\textbf{Closed- and open-book models.}
Both settings use the same visual backbone and mechanism-decoding interface.
The closed-book model bypasses the external precedent branch and directly
decodes from the current query representation $H_q^t$.
The open-book model additionally applies precedent retrieval, cross-image
transfer, and gated fusion to obtain $\widetilde{H}_q^t$ before decoding.
Structured electron-edit serialization and execution are described separately
in the following section.

\textbf{OCSR-based structured retrieval baseline.}
For the structured retrieval baseline, we preprocess the external precedent bank with OpenChemIE~\citep{fan2024openchemie}, converting each literature figure into structured representations of the depicted molecular structures and reaction connectivity. The resulting symbolic outputs are used to construct the structured retrieval index, while figures for which OpenChemIE fails to produce a valid extraction are excluded. This pipeline provides an OCSR-based alternative to directly indexing the precedent bank in its native visual form.

\section{Molecular Image Representation Details} 
Standard molecular renderings~\cite{joung2025electron} often suffer from visual overlaps (e.g., overlapping groups and occluded bonds) that hinder accurate reasoning. To create clear and reliable visual inputs, we apply three rendering optimizations:

\textbf{Implicit hydrogen omission.}
We omit predictable implicit hydrogens to substantially reduce visual clutter while preserving their mechanistic information. When an implicit hydrogen participates in an electron-transfer step, it is represented through the heavy atom to which it is attached rather than by a separate visual index. For example, if an omitted hydrogen is bonded to atom \texttt{C:137}, we denote it as \texttt{H:(C:137)}, which uniquely identifies the hydrogen through its attachment site. Electron changes involving this hydrogen can therefore still be specified and mapped back to the corresponding heavy atom, so removing its explicit rendering does not discard the information required for mechanism prediction. We implement this using a modified RDKit \texttt{RemoveHs} procedure that removes implicit hydrogens while retaining chemically explicit hydrogen species, such as H--H and H$^+$.

\textbf{Adjacent indexing}: We place atom indices next to the symbols instead of on top of them. This prevents numbers from blocking bonds and keeps local structures clear.
This is implemented by enabling the \textit{addAtomIndices} option when rendering molecular graphs with RDKit.

\textbf{Adaptive layout optimization.}
For complex molecules, we dynamically adjust bond lengths, bond
angles, and fragment placements until all remaining spatial overlaps
are resolved. We formulate this step as a constrained energy
minimization problem,
\begin{equation}
E \;=\; w_{1}E_{\mathrm{overlap}} \;+\; w_{2}E_{\mathrm{cross}}
\;+\; w_{3}E_{\mathrm{bond}} \;+\; w_{4}E_{\mathrm{angle}}
\;+\; w_{5}E_{\mathrm{ring}},
\label{eq:layout-energy}
\end{equation}
where $E_{\mathrm{overlap}}$ is a quadratic penalty on non-bonded
atom pairs within a cutoff $d_{\min}$, $E_{\mathrm{cross}}$ counts
intersections between non-adjacent bond segments,
$E_{\mathrm{bond}}$ and $E_{\mathrm{angle}}$ are harmonic deviations
from ideal bond lengths $l_0$ and from ideal bond angles $\theta_0$
(set per hybridization state, e.g.\ $120^{\circ}$ for $sp^{2}$,
$109.5^{\circ}$ for $sp^{3}$), and $E_{\mathrm{ring}}$ regularizes
ring atoms toward their canonical regular-polygon positions. We set
$w_{1},w_{2} \gg w_{3},w_{4},w_{5}$, treating overlap and bond
crossing as primary objectives while canonical geometry is enforced
as a soft prior.

The optimization proceeds in four stages, releasing degrees of
freedom from coarse to fine.
\emph{(i) Multi-seed initialization.} We generate a pool of candidate
layouts by combining CoordGen with RDKit's stochastic 2D embedder
under multiple random seeds and structural configurations, and select
the lowest-$E$ candidate as the starting point.
\emph{(ii) Rigid-fragment reorientation.} Each ring system and each
substituent fragment is rotated as a rigid body about its connecting
bond over a $30^{\circ}$ grid
$\{0^{\circ},\pm 30^{\circ},\pm 60^{\circ},\pm 90^{\circ},\pm
120^{\circ},\pm 150^{\circ},180^{\circ}\}$ and translated along $12$
radial offset directions, accepting the configuration that minimizes
$E$. This stage resolves the majority of inter-fragment crossings
without altering any internal geometry.
\emph{(iii) Branch-wise local relaxation.} For every atom of degree
$\geq 3$, we refine its outgoing single-bond branches in two passes:
bond angles are first perturbed within $\pm 15^{\circ}$ of $\theta_0$
in $5^{\circ}$ steps, then bond lengths are scaled within
$[0.9, 1.4]\,l_0$, with the range escalated to
$[1.4, 1.8]\,l_0$ only when the local overlap remains unresolved.
Double, triple, and ring bonds are held rigid throughout to preserve
canonical chemical geometry.
\emph{(iv) Coordinated escape from local minima.} Greedy single-branch
updates can stagnate when overlaps require two or more branches to
move in concert. Whenever no improving move is found for two
consecutive rounds, we jointly search the top $k{=}3$ most crowded
branches over their combined offset space, and additionally apply
simulated-annealing-style random perturbations accepted with
probability $\min(1, \exp(-\Delta E / T))$, with the temperature $T$
decayed geometrically across rounds.

The procedure terminates once no non-bonded pair violates $d_{\min}$
and no bonds cross, or when a patience budget of $5$ consecutive
non-improving rounds is exhausted.

\begin{figure}[t]
    \centering
    \includegraphics[width=1.0\textwidth]{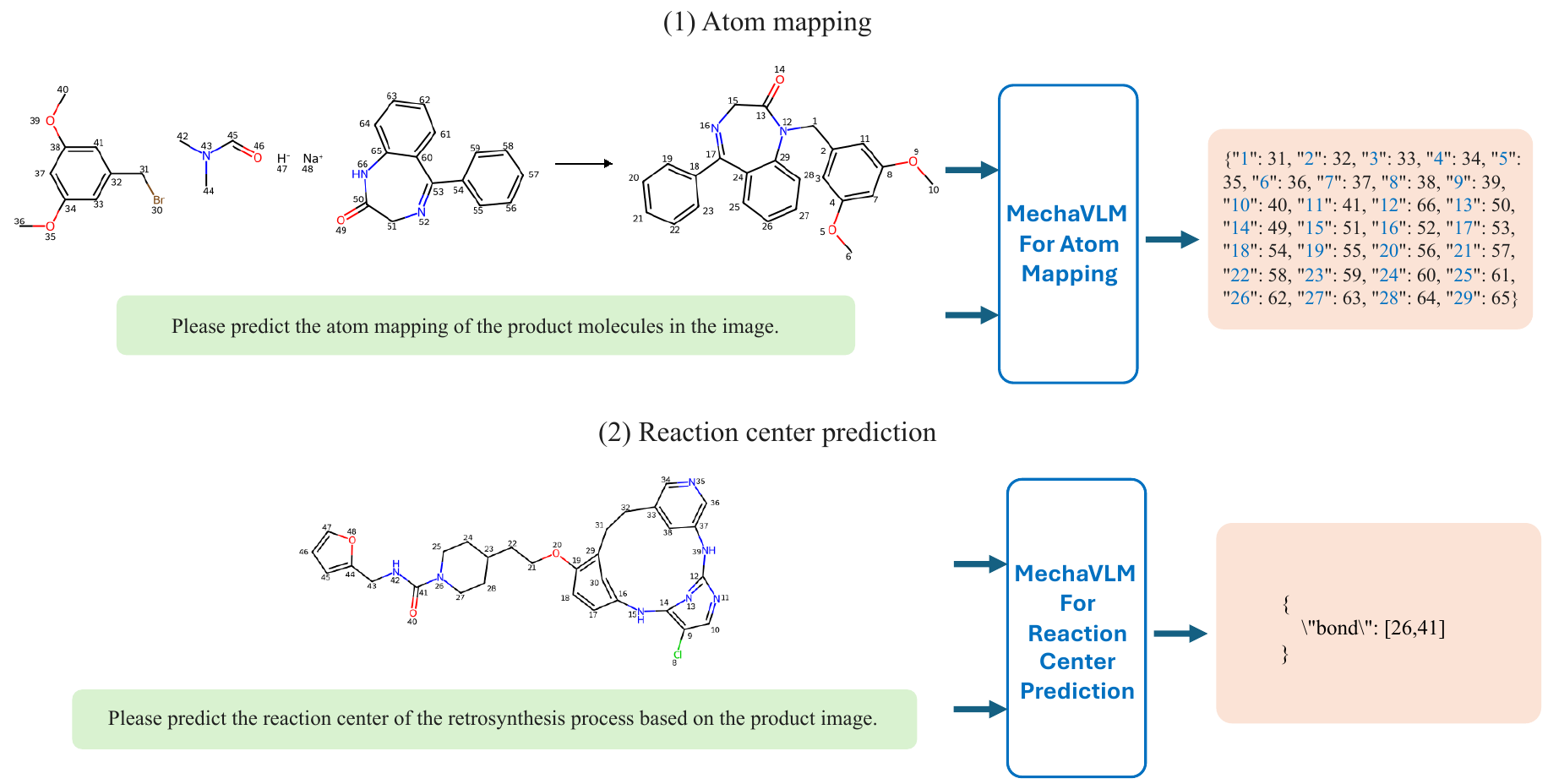}
    \vspace{-2em}
    \caption{
    Example of input/output of the image-based atom mapping and reaction center prediction.
    }
    \label{fig:input and output transfer}
    \vspace{-1em}
\end{figure}

\section{Transferable Chemical Representations Beyond Mechanism Prediction}
Beyond its primary objective, we demonstrate that vision-based mechanism prediction provides a highly effective prior for broader reaction-related tasks, specifically atom mapping and reaction center prediction. 
To evaluate this transferability, we render standard molecular inputs into 2D images (see Fig.~\ref{fig:input and output transfer}) to reformulate these tasks into visual reasoning tasks. We then fine-tune the model with and without mechanistic training to explicitly quantify the performance gains.

\textbf{Atom Mapping.} 
We finetune MechaVLM on USPTO-FULL from~\cite{chen2024precise}, and the USPTO-50K test set is created by randomly choosing 2k samples not seen in USPTO-FULL (both the training set of atom mapping and the reaction mechanism prediction).
This guarantees that there is no data leakage in our experimental setting.
In Tab.~\ref{tab:pretrain_ablation}, 
Schneider test set is also the randomly chosen 2k samples from 50k samples in~\cite{schwaller2021extraction}. Jaworski contains annotated 624 samples from ~\cite{schwaller2021extraction}.
These test dataset has no overlap with the training set.
Our mechanistic pre-training yields substantial gains over the from-scratch baseline, enabling our vision-centric framework to surpass specialized tools.

\textbf{Reaction Center Prediction.} 
Tab.~\ref{tab:reaction_center_ablation} further validates this dual benefit for identifying reaction sites on the USPTO-50K dataset. We employ an official split to partition this dataset into training, validation, and test sets. To explicitly prevent data leakage, we implemented a strict reaction-level isolation: all specific reaction instances present in the USPTO-50K test set were completely excluded from the entire MechaVLM training corpus (including the USPTO-FULL dataset). This rigorous isolation ensures that the model is evaluated on entirely unseen molecular structures.

\section{Additional Experiments and Analysis}
\label{app:additional_analysis}

We provide additional analyses complementing Sec.~\ref{sec:ablation}, covering prediction validity and conservation, grounding granularity, retrieval frequency, and scaling with external mechanistic knowledge. Unless otherwise specified, all models are trained on FlowER.

\textbf{Validity and conservation.}
As shown in Tab.~\ref{tab:smiles_valid}, MechaVLM produces highly valid and chemically consistent predictions, achieving 95.12\% validity and conservation across atoms, protons, and electrons. This substantially exceeds sequence-based baselines and slightly surpasses FlowER, indicating that image-based mechanism prediction does not sacrifice chemical validity despite operating directly from molecular diagrams.

\begin{table}[!htbp]
\centering
\footnotesize
\setlength{\tabcolsep}{10pt}
\renewcommand{\arraystretch}{0.95}

\caption{Comparison of prediction validity and conservation (\%).}
\begin{tabular}{lcccc}
\toprule[1.5pt]
& & \multicolumn{3}{c}{\textbf{Conservation}} \\
\cmidrule(l){3-5}
\textbf{Method} & \textbf{Validity} & Atom & Proton & Electron \\
\midrule[1pt]

Graph2SMILES~\cite{tu2022permutation}
& 76.31 & 30.72 & 18.67 & 17.19 \\

Graph2SMILES+H~\cite{tu2022permutation}
& 78.80 & 27.66 & 19.73 & 19.02 \\

Molecular Transformer~\cite{schwaller2019molecular}
& 70.16 & 39.11 & 33.69 & 33.04 \\

FlowER~\cite{joung2025electron}
& 94.94 & 94.94 & 94.94 & 94.94 \\

\midrule
\textbf{MechaVLM}
& \textbf{95.12} & \textbf{95.12} & \textbf{95.12} & \textbf{95.12} \\

\bottomrule[1.5pt]
\end{tabular}
\label{tab:smiles_valid}
\end{table}

\textbf{Effect of Grounding Granularity.}
We further examine whether the transfer gains from chemically grounded pretraining arise only from atom-level recognition or from jointly modeling chemical units at multiple scales. Starting from the same backbone, we progressively add atom-, bond-, and functional-group-level supervision during Stage-I pretraining while keeping the subsequent mechanism training unchanged. As shown in Tab.~\ref{tab:grounding_granularity}, atom-level grounding provides the largest initial improvement by anchoring visual features to elemental identity and formal charge. Adding bond supervision further improves transfer by capturing local connectivity and bond order, while functional-group grounding provides additional gains by encoding higher-level recurring chemical environments. Notably, these improvements are much larger on the three OOD benchmarks than on FlowER, suggesting that multiscale chemical grounding mainly improves the transferability of learned representations rather than source-domain fitting.

\begin{table}[!htbp]
\centering
\small
\caption{
Effect of chemical grounding granularity (\%).
All variants use the same Stage-I training data and subsequent mechanism
training protocol.
Only Top-1 Step/Pathway accuracy is reported.
}
\label{tab:grounding_granularity}
\setlength{\tabcolsep}{4.5pt}
\renewcommand{\arraystretch}{1.08}
\resizebox{\linewidth}{!}{
\begin{tabular}{@{}lcccccccc@{}}
\toprule
&
\multicolumn{2}{c}{\textbf{FlowER (ID)}}
&
\multicolumn{2}{c}{\textbf{ReactMech}}
&
\multicolumn{2}{c}{\textbf{FukuyamaBench}}
&
\multicolumn{2}{c}{\textbf{MechBench}} \\
\cmidrule(lr){2-3}
\cmidrule(lr){4-5}
\cmidrule(lr){6-7}
\cmidrule(lr){8-9}

\textbf{Grounding supervision}
& Step & Path.
& Step & Path.
& Step & Path.
& Step & Path. \\
\midrule

None
& 88.41 & 91.96
& 52.32 & 32.47
& 15.87 & 1.57
& 6.82 & 0.60 \\

Atom
& 88.86 & 92.51
& 57.18 & 37.06
& 17.38 & 2.19
& 7.54 & 0.78 \\

Atom + Bond
& 89.31 & 93.08
& 60.74 & 42.26
& 18.63 & 3.13
& 8.16 & 0.96 \\

Atom + Bond + Functional Group
& \textbf{89.63} & \textbf{93.42}
& \textbf{62.74} & \textbf{45.58}
& \textbf{19.58} & \textbf{3.76}
& \textbf{8.58} & \textbf{1.10} \\

\bottomrule
\end{tabular}
}
\end{table}

\textbf{Reaction-Level vs.\ Step-Wise Retrieval.}
We compare our reaction-level retrieval with step-wise re-retrieval, while keeping all other components unchanged. As shown in Tab.~\ref{tab:retrieval_frequency}, fixing the precedent set per reaction consistently performs better across all three OOD benchmarks. Re-retrieval from evolving intermediates can introduce precedent drift, replacing globally relevant pathways with locally similar but less relevant figures. In contrast, MechaVLM preserves reaction-level context while dynamically selecting step-relevant evidence from a fixed precedent set, and reduces retrieval cost from $T$ queries to one for a $T$-step pathway.

\begin{table}[!htbp]
\centering
\small
\caption{
Effect of precedent retrieval frequency (\%).
}
\label{tab:retrieval_frequency}
\setlength{\tabcolsep}{5pt}
\renewcommand{\arraystretch}{1.08}
\begin{tabular}{@{}lcccccc@{}}
\toprule
&
\multicolumn{2}{c}{\textbf{ReactMech}}
&
\multicolumn{2}{c}{\textbf{FukuyamaBench}}
&
\multicolumn{2}{c}{\textbf{MechBench}} \\
\cmidrule(lr){2-3}
\cmidrule(lr){4-5}
\cmidrule(lr){6-7}

\textbf{Retrieval strategy}
& Step & Path.
& Step & Path.
& Step & Path. \\
\midrule

Re-retrieve at every step
& 74.68 & 58.93
& 25.87 & 7.47
& 12.08 & 2.15 \\

\textbf{Reaction-level retrieval (ours)}
& \textbf{76.21} & \textbf{61.52}
& \textbf{27.14} & \textbf{8.15}
& \textbf{12.72} & \textbf{2.52} \\

\bottomrule
\end{tabular}
\end{table}

\textbf{Scaling with External Mechanistic Knowledge.}
We evaluate how open-book transfer scales with the size of the external precedent bank by keeping model parameters fixed and restricting retrieval to nested subsets of the full corpus. As shown in Tab.~\ref{tab:bank_scaling}, performance improves consistently as more precedents become accessible, with the largest gain appearing when moving from closed-book prediction to partial external memory and smaller gains thereafter. This trend suggests that MechaVLM benefits from broader mechanistic coverage rather than a small set of favorable precedents, and that additional external knowledge can improve transfer at inference time without retraining or target-domain supervision.

\begin{table}[!htbp]
\centering
\small
\caption{
Scaling with the external visual precedent bank (\%).
}
\label{tab:bank_scaling}
\setlength{\tabcolsep}{5pt}
\renewcommand{\arraystretch}{1.08}
\begin{tabular}{@{}lccccc@{}}
\toprule
&
&
\multicolumn{2}{c}{\textbf{FukuyamaBench}}
&
\multicolumn{2}{c}{\textbf{MechBench}} \\
\cmidrule(lr){3-4}
\cmidrule(lr){5-6}

\textbf{Bank fraction}
& \textbf{\# Figures}
& Step & Path.
& Step & Path. \\
\midrule

$0\%$   & 0      & 22.28 & 5.96 & 9.59  & 1.56 \\
$25\%$  & 17,596 & 24.29 & 6.58 & 10.74 & 1.88 \\
$50\%$  & 35,192 & 25.74 & 7.52 & 11.58 & 2.15 \\
$100\%$ & 70,384 & \textbf{27.14} & \textbf{8.15}
                    & \textbf{12.72} & \textbf{2.52} \\

\bottomrule
\end{tabular}
\end{table}

\begin{figure}[!htbp]
    \centering
    \includegraphics[width=1.0\textwidth]{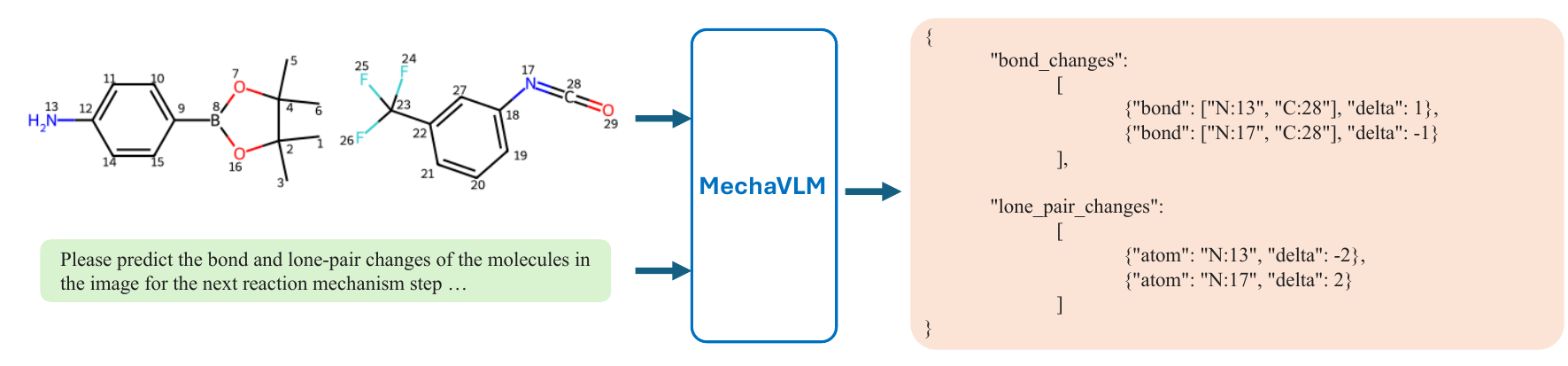}
    \vspace{-2em}
    \caption{
    Example of input and output format of MechaVLM in closed-book setting.
    }
    \vspace{-1em}
    \label{fig:input and output}
\end{figure}

\section{Input and Output Format of MechaVLM}
Fig.~\ref{fig:input and output} illustrates an example of complete input and output of MechaVLM in closed-book setting.

\section{More Visualization of OOD prediction}
Fig.~\ref{fig:input and output} shows more examples of MechaVLM's OOD prediction in closed-book/open-book setting.

\begin{figure}[!htbp]
    \centering
    \includegraphics[width=1.0\textwidth]{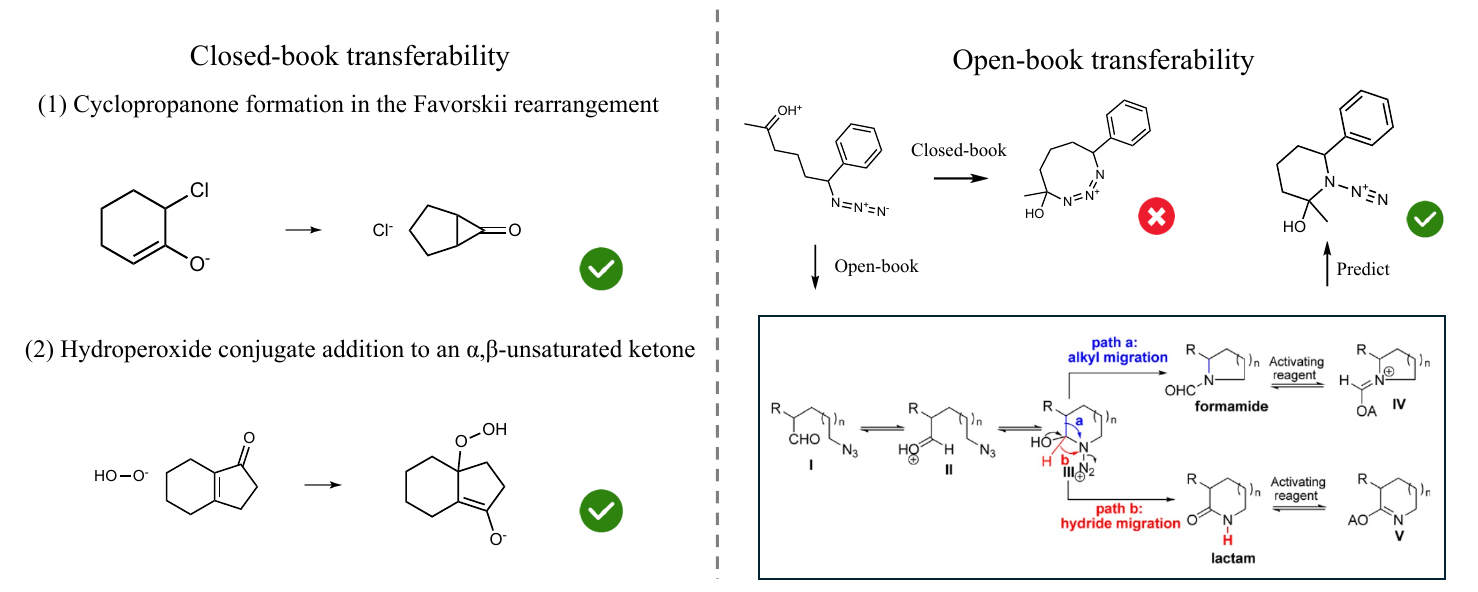}
    \vspace{-1em}
    \caption{
    Example of input and output format of MechaVLM in closed-book setting.
    }
    \label{fig:OOD examples suppl}
    \vspace{-1em}
\end{figure}

\section{Model Details}

MechaVLM adopts Qwen2.5-VL-3B as the backbone VLM.
Its main image-encoder and language-model configurations are summarized in
Tab.~\ref{tab:model_details}.

\begin{table}[!htbp]
\footnotesize
\centering
\setlength{\tabcolsep}{8pt}
\renewcommand{\arraystretch}{0.92}
\caption{Architecture details of the Qwen2.5-VL-3B backbone.}
\label{tab:model_details}
\begin{tabular}{lc}
\toprule
\textbf{Component} & \textbf{Qwen2.5-VL-3B} \\
\midrule
\multicolumn{2}{l}{\emph{Image Encoder}} \\
Architecture & Vision Transformer \\
Layers & 32 \\
Hidden Size & 1280 \\
Attention Heads & 16 \\
Patch Size & 14 \\
\midrule
\multicolumn{2}{l}{\emph{Language Model}} \\
Architecture & Transformer LLM \\
Layers & 36 \\
Hidden Size & 2048 \\
KV Heads & 2 \\
Head Size & 128 \\
Positional Encoding & Rotary \\
\bottomrule
\end{tabular}
\end{table}

\end{document}